\documentclass[prc,preprint,showpacs,
                floatfix,amsfonts,nofootinbib]{revtex4}
\usepackage{latexsym}
\usepackage{graphicx}
\usepackage{amssymb}
\usepackage{amsmath}
\usepackage{amscd}

\newcommand{\ugamma}[1]{\gamma^{#1}}

\newcommand{\usigma}[1]{\sigma^{#1}}

\newcommand{\ket}[1]{\vert#1\rangle}
\newcommand{\bra}[1]{\langle#1\vert}

\newcommand{\psibar}[1]{\overline{#1}}
\newcommand{\half}{\ensuremath{\frac{1}{2}}}

\newcommand{\slashed}[1]{\not\!#1}
\newcommand{\lag}{\mathcal{L}}

\newcommand{\ucpartial}[1]{\widetilde{\partial}^{#1}}
\newcommand{\dcpartial}[1]{\widetilde{\partial}_{#1}}
\newcommand{\mn}{\mu\nu}

\newcommand{\modular}[2]{\vert \vec{#1}_{#2}\vert}
\newcommand{\starobject}[1]{#1^{\ast}}
\def\Tr{\mathop{\rm Tr}\nolimits}
\def\Mstar{M^{\ast}}
\def\mstar{m^{\ast}}

\begin{document}
\title{Incoherent Neutrinoproduction of Photons and Pions in a Chiral 
            Effective Field Theory for Nuclei}
            
\author{Xilin Zhang}\email{xilzhang@indiana.edu}
\author{Brian D. Serot} \thanks{Deceased.}
\affiliation{Department of Physics and Center for Exploration of
             Energy and Matter\\
             Indiana University, Bloomington, IN\ \ 47405}

\date{\today}

\begin{abstract}

We study the incoherent neutrinoproduction of 
photons and pions with neutrino energy $E_{\nu} \leqslant 0.5 \ \mathrm{GeV}$. 
These processes are relevant to the background analysis in 
neutrino-oscillation experiments [for example, MiniBooNE; A. A. Aquilar-Arevalo \textit{et al.} (MiniBooNE Collaboration), 
Phys.\ Rev.\ Lett.\ {\bf 100}, 032301 (2008)]. 
The calculations are carried out using a Lorentz-covariant effective 
field theory (EFT),
which contains nucleons, pions, the Delta (1232) ($\Delta$), 
isoscalar scalar ($\sigma$) and 
vector ($\omega$) fields, and isovector vector ($\rho$) fields, and has
$\mathrm{SU(2)}_{\mathrm{L}} \otimes \mathrm{SU(2)}_{\mathrm{R}}$ chiral symmetry realized nonlinearly. 
The contributions of one-body currents are studied in the local Fermi gas approximation.  
The current form factors are generated by meson dominance in the 
EFT Lagrangian. The conservation of the vector current and 
the partial conservation of the axial current are satisfied automatically, which 
is crucial for photon production. 
The $\Delta$ dynamics in nuclei, as a key component in the study, is explored.  
Introduced $\Delta$-meson couplings explain the $\Delta$ spin-orbit coupling in nuclei, and 
this leads to interesting constraints on the theory. Meanwhile a phenomenological 
approach is applied to parametrize the $\Delta$ width.   
To benchmark our approximations, we calculate the differential cross sections 
for quasi-elastic scattering and incoherent electroproduction of pions 
without a final state interaction (FSI). 
The FSI can be ignored for photon production. 
\end{abstract}

\smallskip

\maketitle

\section{Introduction}

This paper is a continuing work of \cite{bookchapter,1stpaper}, 
focusing on neutrinoproduction of photons and pions from nuclei with neutrino energy 
$E_{\nu} \leqslant 0.5 \ \mathrm{GeV}$. 
In Refs.~\cite{bookchapter,1stpaper}, we introduced the $\Delta$ resonance as a manifest 
degrees of freedom to the effective field theory (EFT), known as \emph{quantum hadrodynamics} or QHD 
\cite{SW86,SW97,Furnstahl9798,FSp00,FSL00,EvRev00,LNP641,EMQHD07}. (The
motivation for this EFT and some calculated results are discussed in
Refs.~\cite{SW97,Furnstahl9798,HUERTAS02,HUERTASwk,HUERTAS04,MCINTIRE04,%
MCINTIRE05,JDW04,MCINTIRE07,HU07,MCINTIRE08,BDS10}.)
To calibrate the reaction mechanism on the nucleon level, 
we studied the productions from nucleons \cite{1stpaper}. 
The calculations are motivated by the fact that the 
neutrinoproductions of $\pi^0$ and photons from nuclei 
(and nucleons) are potential backgrounds in  
neutrino-oscillation experiments (e.g., MiniBooNE \cite{MiniBN2007,MiniBN2009,MiniBN2010}). 
Currently, it is still a question whether the neutral current 
(NC) photon production might explain the excess events seen at 
low \emph{reconstructed} neutrino energies, 
which the MicroBooNE experiment plans to answer \cite{MicroBN2011}.  
Moreover, the authors of Refs.~\cite{Harvey07, Harvey08, Hill10, Gershtein81} point 
out the possible role of anomalous interaction vertices involving 
$\omega (\rho)$, $Z$, and the photon in NC photon 
production. 
So it is necessary to calculate the cross sections for these processes. 
Here by using the QHD EFT, we  
study incoherent production, 
in which the nucleus is excited.
Coherent production with the nucleus 
being intact is a topic of future work. \footnote{Recently, a 
unified framework for handling both coherent and incoherent production has been proposed 
in \cite{Martini09}.}
We will discuss the power-counting 
\footnote{%
In an EFT, there are an infinite number of interaction terms 
allowed by various constraints. To organize them, 
we can associate 
power-counting to each vertex and diagram. The calculation 
can be done in a perturbative way by summing 
diagrams up to some particular power $\nu$. 
See Refs.\cite{Furnstahl9798,
MCINTIRE07,HU07,MCINTIRE08,BDS10, bookchapter,1stpaper} for detailed discussions 
about power-counting in QHD EFT.%
} 
of the calculations through which we will show that the contributions of
the anomalous interactions are small in the 
incoherent NC production of photons
(where they contribute at next-to-next-to-leading-order).
To benchmark the approximation scheme, 
we study electron scattering in both quasi-elastic 
and pion production channels. 

There have been several 
experiments measuring the weak response of nuclei across the 
quasi-elastic 
region to the $\Delta$ excitation peak. In most experiments 
\cite{k2k05plb, k2k05prl, k2k06prd, k2k08prd, Sciboone08prd, 
Miniboone08prl, Miniboone08plb, Miniboone09prl}, which 
have ${}^{12}C$ and ${}^{16}O$ as the primary target nuclei, 
the mean energy of the beam is 
around $ 1 \ \mathrm{GeV}$. As emphasized in 
\cite{1stpaper}, we expect our theory to work up to 
$0.5$ GeV, so we do not rely on these experiments to 
constrain the theory at this stage. 
On the theoretical side, much work has been done (e.g., in 
\cite{Martini09, Butkevich08, Butkevich09, Singh92, Singh93, 
Meucci04_1, Meucci04_2, Nieves04, Martinez06, 
GiBUU2006cc, GiBUU2006nc, GiBUU2009, Kartavtsev06, 
Praet09, GiBUUprc79_057601, Singh98, Sato03, 
Szczerbinska07, Amaro05, Caballero05, 
Amaro07, Martini07, Amaro07_prc, Ivanov08}). Most of these papers are based on the global or 
local Fermi gas approximation and include contributions from one-body currents, 
with improved treatment for final-state interaction (FSI) and 
$\Delta$ dynamics in the medium. The same approach has also
been applied in electron scattering (e.g., in \cite{Gil97}). 
In \cite{Amaro05, Caballero05, 
Amaro07, Martini07, Amaro07_prc, Ivanov08}, scaling approaches are 
used to address quasi-elastic scattering. 
Moreover, the contribution from two-body currents was studied nonrelativistically,
for example, in \cite{Alberico84}.
In most of these calculations, the $\Delta$  
dynamics in nuclei is based on the work of \cite{oset87}, 
in which the $\Delta$ self-energy has been studied 
using a nonrelativistic model. 
Parallel to the nonrelativistic studies, some work 
has been initiated in the relativistic framework, QHD EFT, 
using the local Fermi gas (LFG) approximation and including 
one-body currents \cite{Rosenfelder80,wehrgerger89, 
wehrgerger90, wehrgerger92, wehrgerger93}. 
The two-body current 
was investigated relativistically in \cite{Dekker94,DePace03}. These works mainly 
focus on electron scattering. 
But the handling of the $\Delta$ resonance in these papers is somewhat 
phenomenological. Moreover, in 
both nonrelativistic and relativistic studies, 
photon production is rarely investigated. 

In this paper, we also apply the LFG approximation 
\cite{Rosenfelder80} to study the one-body current 
contribution. As shown in \cite{bookchapter,1stpaper}, 
we make use of meson dominance to generate form factors 
for various currents. Because of the built in symmetries 
in the Lagrangian, conservation of vector current and the partial conservation of axial 
current are satisfied. These properties are well preserved 
in the LFG approximation. Especially for 
photon production, vector current conservation
is crucial. 
The $\Delta$ dynamics, as a key component 
in this work, is explored to some extent. 
We introduce interactions between $\Delta$ and 
non-Goldstone meson fields to generate the spin-orbit 
(S-L) coupling that has been introduced in 
phenomenological models 
\cite{horikawa80, Nakamura10}. On the other hand, phenomenological 
knowledge about S-L coupling puts constraints on these
couplings. Moreover, the $\Delta$ decay width 
increases in the nucleus, because more decay channels are opened up and 
this effect overcomes the reduction of pion decay phase 
space. Here we follow 
the phenomenological studies and separate the width to the
pion decay width and anything else parametrized by 
the imaginary part of the $\Delta$ spreading potential. 
As a result of opening new decay channels, the flux having excited a 
$\Delta$ resonance can be transferred to channels that do not involve pion or photon production. Moreover, Pauli blocking can reduce the pion and photon production cross section further, 
because of the reduction of the final particle's phase space. 
\footnote{The binding effect should be important when the neutrino energy 
is close to threshold, where the simple approximations used here are not feasible. 
But this is clearly not important around 0.5 GeV.} 
In this paper, we explore how both $\Delta$ and 
nonresonant contributions are reduced compared to those 
in free nucleon scattering.
However, we do not include FSI effects for pions and knocked out nucleons. The simple treatment can be 
found in \cite{Adler79, paschos00}, while the complete 
treatment is implemented in various event generators of 
experiments (e.g., NUANCE \cite{nuance}),  
and the GiBUU model \cite{GiBUU2006cc, GiBUU2006nc}. Hence we only 
compare our predictions with the output of NUANCE without FSI. 
\footnote{The predictions from NUANCE shown throughout this paper are obtained from the NUANCE v3 event generator \cite{nuance}. Multiple resonances are considered in NUANCE, but the $\Delta$ dominates. The axial mass $M_{A}^{\pi}=1.10\pm0.27$ GeV is used which is the same as that used by the MiniBooNE experiment for their \emph{baseline} calculations \cite{MBCCQE}. However, the actual backgrounds used in their final analyses were scaled to data in a separate exercise.}

The paper is organized as follows. In 
Sec.~\ref{sec:quasi}, we first discuss the LFG
approximation and then apply it to electron quasi-elastic 
scattering, which serves as a benchmark. 
In Sec.~\ref{sec:pionprod}, the calculation 
scheme for pion (photon) production 
is briefly introduced. Then the $\Delta$ dynamics is studied with emphasis on the 
connection between $\Delta$-meson interactions and S-L coupling. 
The modification of the $\Delta$ width is also discussed. 
After that, electron scattering at the $\Delta$ peak
is studied, and results are compared with data 
with explanation of the missing strength. The cross sections 
of neutrinoproduction of pions are also shown and compared 
to NUANCE's output. Sec.~\ref{sec:photonprod} is dedicated 
to NC photon production. Finally Sec.~\ref{sec:sum} contains 
a short summary.
In the appendices, we show detailed kinematic analyses for 
both quasi-elastic scattering and pion production.

\section{Quasi-elastic scattering in the LFG approximation} \label{sec:quasi}

This section serves as an illustration of the LFG
approximation used for quasi-elastic scattering and 
for photon and pion production.
(See Ref.~\cite{1stpaper} for discussion on 
the free nucleon interaction amplitude in all these processes.)
Here we make use of the mean-field approximation 
to calculate the nuclear ground state. The relevant leading order 
Lagrangian is
\begin{eqnarray}
\lag = \psibar{N} \left[ i\ugamma{\mu}\left(\dcpartial{\mu}
                        + ig_{\rho}\rho_{\mu}
                        + ig_{v}V_{\mu}\right)
                        - M
                        + g_{s}\phi
                   \right]N  
\end{eqnarray}
(where the full Lagrangian can be found in \cite{bookchapter,Furnstahl9798} for example).
The mean-field approximation is presented simply as 
follows. Inside nuclear matter, 
vector $\rho^{3\mu}$ and $V^{\mu}$, and scalar $\phi$ fields 
develop nonzero expectation values. 
In the laboratory frame of the matter,
only two fields ($\phi$ and $V^{0}$) have nonzero 
values (but in the isospin asymmetric case, 
$\rho^{0}$ can also develop a nonzero value). 
As a result, the nucleon's mass is modified:
$\Mstar=M-g_{s}\langle\phi\rangle$. At the lowest order, 
the spectrum of nucleons is 
$E(\vec{p})=\sqrt{\vec{p}^{2}+M^{\ast 2}}+g_{v}\langle V^{0}\rangle$. 
Inside a finite nucleus, due to different boundary 
conditions, the mean-field expectation value is space dependent 
and can be calculated numerically. By using this approximation, 
we can calculate the bulk properties of the nucleus, the details of which 
can be found in Ref.~\cite{Furnstahl9798} for example. 

\begin{figure}
\centering
\includegraphics[scale=0.7, angle=-90]{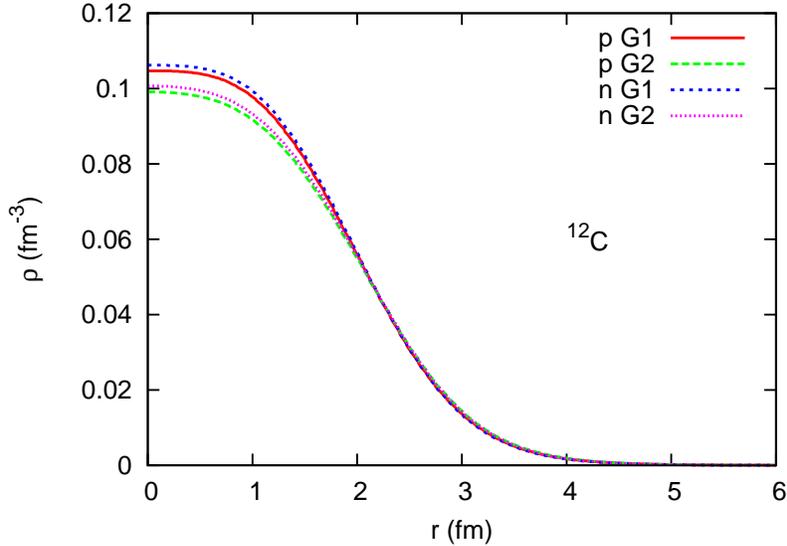}
\caption{(Color online) Proton and neutron density in ${}^{12}C$ with 
         G1 and G2 parameter sets.}
\label{fig:c12density}
\end{figure}

\begin{figure}
\centering
\includegraphics[scale=0.7, angle=-90]{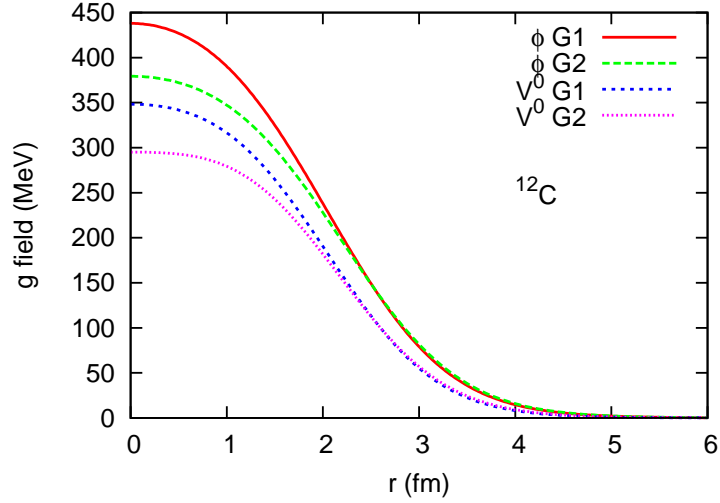}
\caption{(Color online) $\langle g_{s} \phi \rangle$ and $\langle g_{v} V^{0}\rangle $ 
         in ${}^{12}C$ with G1 and G2 parameter sets.}
\label{fig:c12fields}
\end{figure}

Following \cite{Furnstahl9798}, 
we calculate the local density $\rho_{p/n}(\vec{r})$ and field expectation value
in ${}^{12}C$ (the major nucleus in the MiniBooNE's detector). 
Figs.~\ref{fig:c12density} and \ref{fig:c12fields} 
show the results based on $G1$ and $G2$ 
parameter sets in \cite{Furnstahl9798}. We will explore 
the difference due to the two sets in electron quasi-elastic scattering. 

\begin{figure}
\includegraphics[scale=0.9]{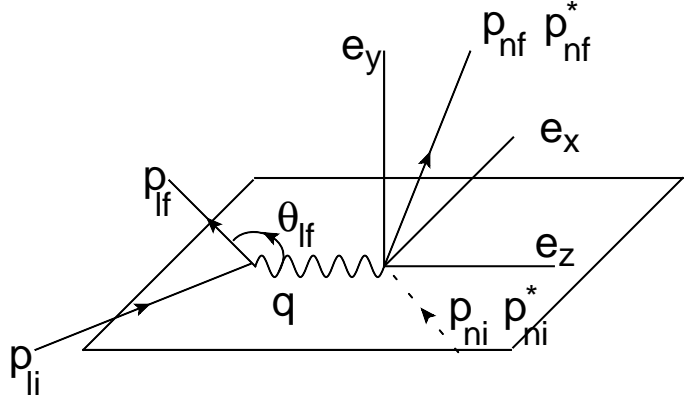}
\caption{The kinematics in the laboratory frame of the nucleus for a lepton 
         interacting with one nucleon inside the nucleus.}
\label{fig:nucleiLabframeconfiguration}
\end{figure}

To calculate the electroweak response of nuclei, 
we use the LFG approximation. 
This approach has been applied in 
\cite{Rosenfelder80} to study electron quasi-elastic 
scattering. First, by assuming the impulse 
approximation (IA), the interaction happens every 
time between probe and each individual nucleon 
This only holds when the 
transferred momentum is high enough that the interference 
between different nucleons is reduced due to the big recoil. 
Second, the response of the nucleus is the incoherent
sum of the response of the fermion gas in different regions.
This works when the probe's wave length 
is small enough compared to a characteristic length scale of 
the nucleus density profile. The discussion can be 
summarized in the following equation:
\begin{eqnarray}
\sigma &=& \int dV \frac{1}{2 p_{li}^{0}} 
           \int \frac{d^{3}{\vec{p}_{nf}}^{\ast}}{(2\pi)^{3}2p_{nf}^{\ast 0}} 
           \frac{d^{3}\vec{p}_{lf}}{(2\pi)^{3} 2 p_{lf}^{0}} 
           \frac{d^{3}{\vec{p}_{ni}}^{\ast}}{(2\pi)^{3} 2 p_{ni}^{\ast 0}} 
           (2\pi)^{4} \delta^{4}(q+\starobject{p_{ni}}-\starobject{p_{nf}}) 
           \underset{s_{f},s_{i}}{\sum} \vert M_{fi} \vert^{2} \ . \label{eqn:quasiTcrosssection}
\end{eqnarray}
In this equation, $p_{li}$ and $p_{lf}$ are the 
incoming and outgoing lepton momenta, respectively, 
$q\equiv p_{li}-p_{lf}$ is the momentum transfer, 
$p_{ni}$ $s_{i}$ and $p_{nf}$ $s_{f}$ are the scattered 
nucleon's initial and final momenta and spin projection, and $M_{fi}$ is the 
one-body interaction amplitude. The kinematic configuration is 
shown in Fig.~\ref{fig:nucleiLabframeconfiguration}. The integration over 
initial and final nucleon momenta depends on the 
space dependent Fermi momentum. A detailed discussion about 
this equation can be found in Appendix~\ref{app:quasikinematics}. 

The interaction amplitude $M_{fi}$ in Eq.~(\ref{eqn:quasiTcrosssection})
can be expressed in terms of 
various current matrix elements (where $V^{i}_{\mu}$, $A^{i}_{\mu}$, and $J_{B\mu}$ 
are, respectively, the isovector vector current, the isovector axial-vector current and the baryon current, 
$i=\pm 1, 0$ \cite{bookchapter,1stpaper}).
For electron quasi-elastic scattering,
\begin{eqnarray}
M_{fi} &=& \frac{e^{2}}{q^{2}} \langle J^{(lep)}_{EM \ \mu} \rangle  
           \langle J_{EM}^{(had) \ \mu} \rangle  \ , \notag \\[5pt]
\langle J_{EM}^{(had)  \mu} \rangle 
     &\equiv& \bra{N, B} V^{0\mu}+\half J_{B}^{\mu}\ket{N, A} \ ,
     \label{eqn:matrixelementEMcurrent}  
\end{eqnarray}
where $A$ and $B$ in the state are 
nucleon isospin.
For charged current (CC) quasi-elastic scattering, 
\begin{eqnarray}
M_{fi} &=& 4 \sqrt{2} G_{F} V_{ud} \langle J_{L i \mu}^{(lep)} \rangle  
           \langle J_{L}^{(had) i \mu} \rangle  \notag  \ , \\[5pt]
\langle J_{L}^{(had) i \mu} \rangle 
       &\equiv& \bra{N, B} \half (V^{i \mu}+A^{i \mu})\ket{N, A} \ ,
       \label{eqn:matrixelementCCcurrent}
\end{eqnarray}
where $i=\pm 1$, $G_{F}$ is the Fermi constant 
and $V_{ud}$ is the $u$ and $d$ quark mixing element in the Cabibbo-Kobayashi-Maskawa (CKM) matrix.
For NC quasi-elastic scattering,
\begin{eqnarray}
M_{fi} &=& 4 \sqrt{2} G_{F} \langle J_{NC \mu}^{(lep)} \rangle 
           \langle J_{NC}^{(had) \mu} \rangle \ ,  \notag   \\[5pt]
\langle J_{NC}^{(had) \mu} \rangle 
      &\equiv& \bra{N, B} J_{L}^{0\mu} - \sin^{2}\theta_{w}J_{EM}^{\mu} 
               \ket{N, A} \ , \label{eqn:matrixelementNCcurrent}
\end{eqnarray} 
where $\theta_{w}$ is the weak mixing angle.
The electroweak currents of leptons are well known, 
and 
$\bra{N, B} V^{i}_{\mu} (J_{\mu}^{B}, A^{i}_{\mu}) \ket{N, A}$ 
can be found in \cite{1stpaper}. But 
we need to include the nucleon spectrum modification to the results of \cite{1stpaper}, 
which is straightforward to complete in the LFG approximation.  

A short discussion on FSI is in order here. The picture is the following: the 
interaction channels are opened in the initial interacting 
vertex, and then these channels would couple to each other when 
particles are traveling through the nucleus. The flux 
among all the initial channels are redistributed due to FSI. 
This picture is adopted in the GiBUU model calculations  
for example for CC and NC processes \cite{GiBUU2006cc} \cite{GiBUU2006nc}. 
From conservation of probability, assuming the picture mentioned 
above is valid, we should expect the sum 
of these channels in the initial vertex to match the inclusive data. 
Moreover, Coulomb distortion of the electron is not 
included in this calculation.

\begin{figure}
\includegraphics[scale=0.7,angle=-90]{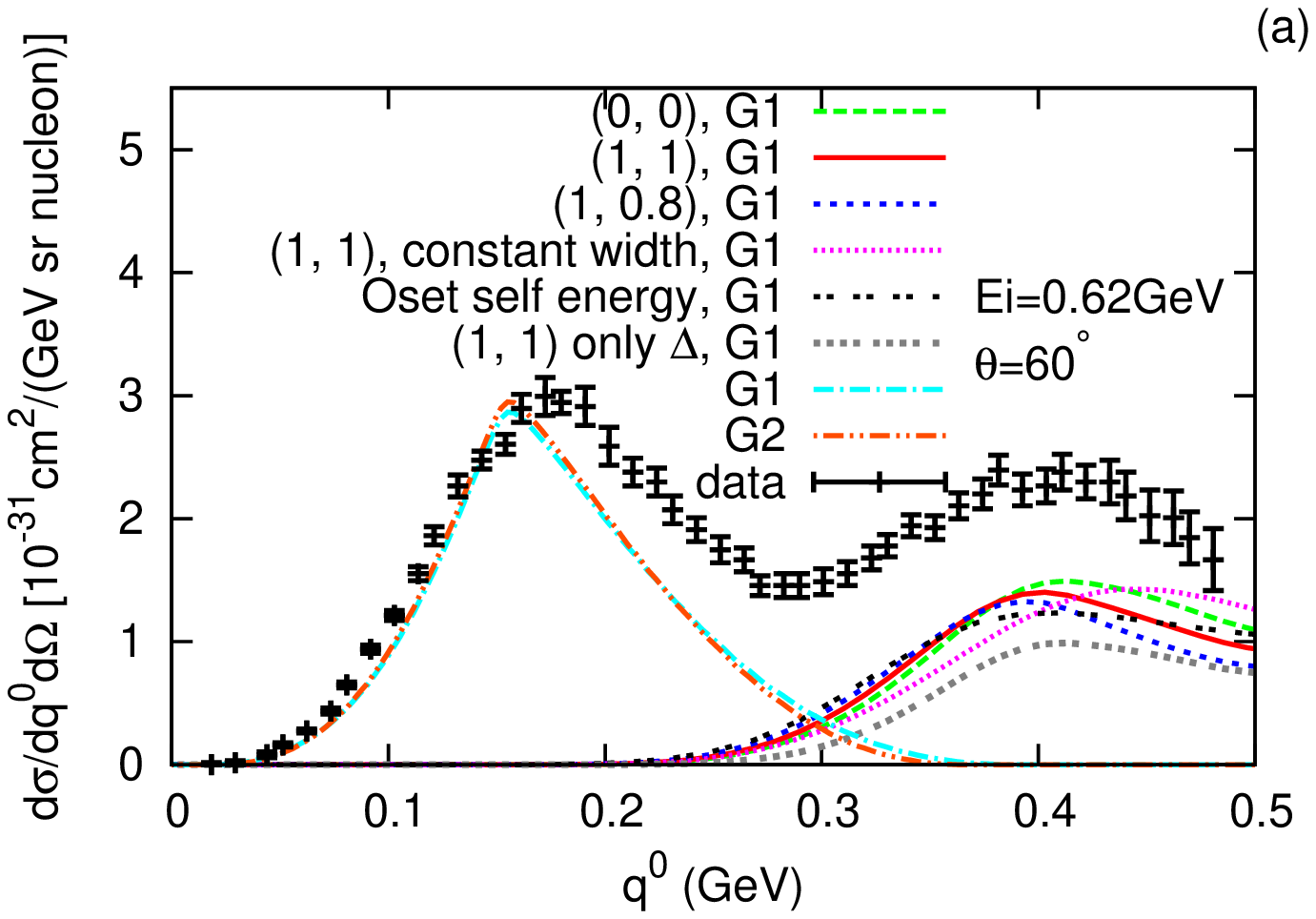}
\includegraphics[scale=0.7,angle=-90]{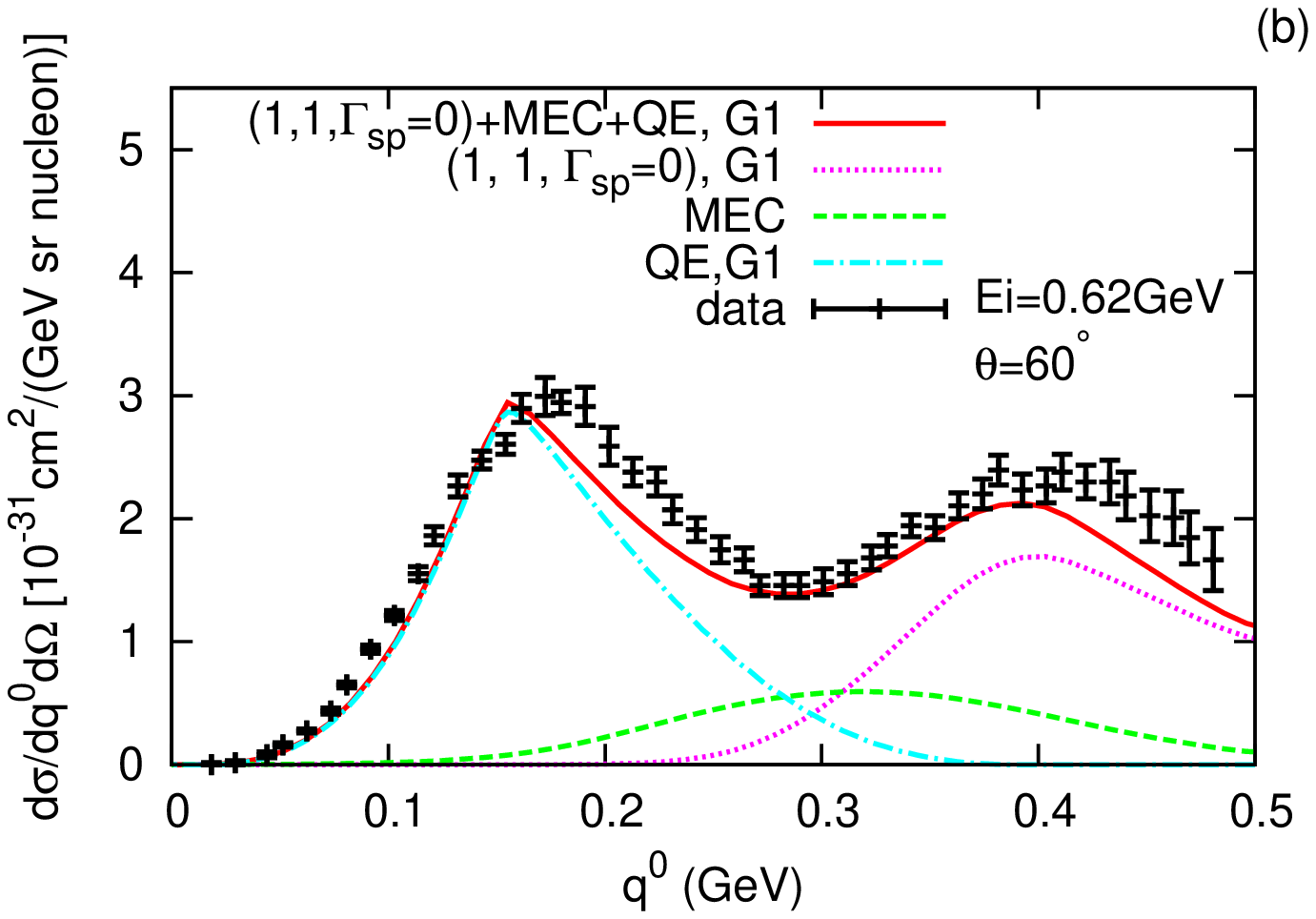}
\caption{(Color online) Inclusive data for differential cross section of 
         electron scattering off ${}^{12}C$. The incoming electron 
         energy is $E_{i}=0.62 \ \mathrm{GeV}$, 
         and the scattering angle is $\theta_{lf}=60^{\circ}$. The kinematics is measured in the laboratory frame. 
         The data are from Ref.~\cite{Barreau83}. Explanations of different plots can be 
         found in the text.
        }
\label{fig:emscattering_0.62_60}
\end{figure}

\begin{figure}
\includegraphics[scale=0.7,angle=-90]{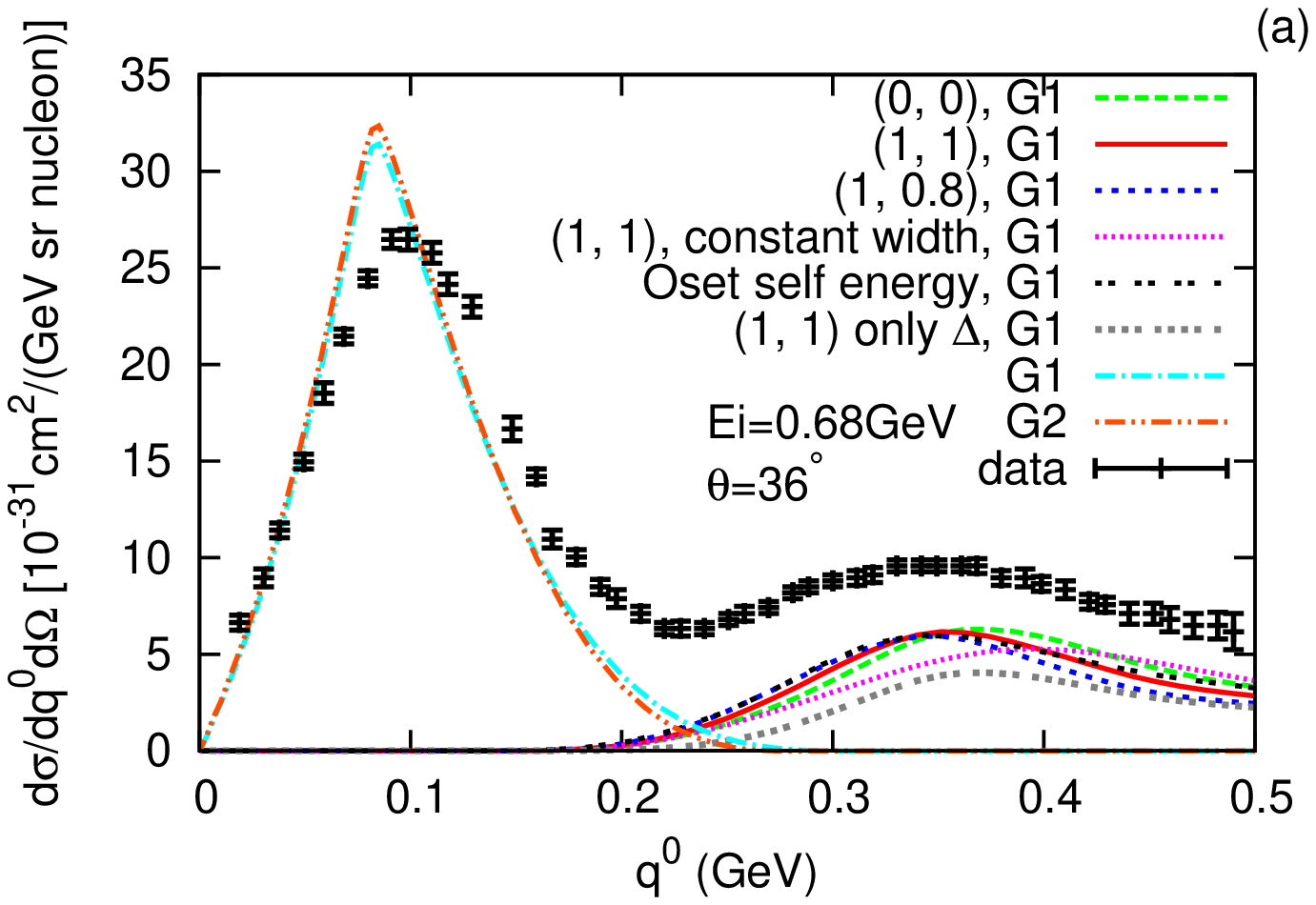}
\includegraphics[scale=0.7,angle=-90]{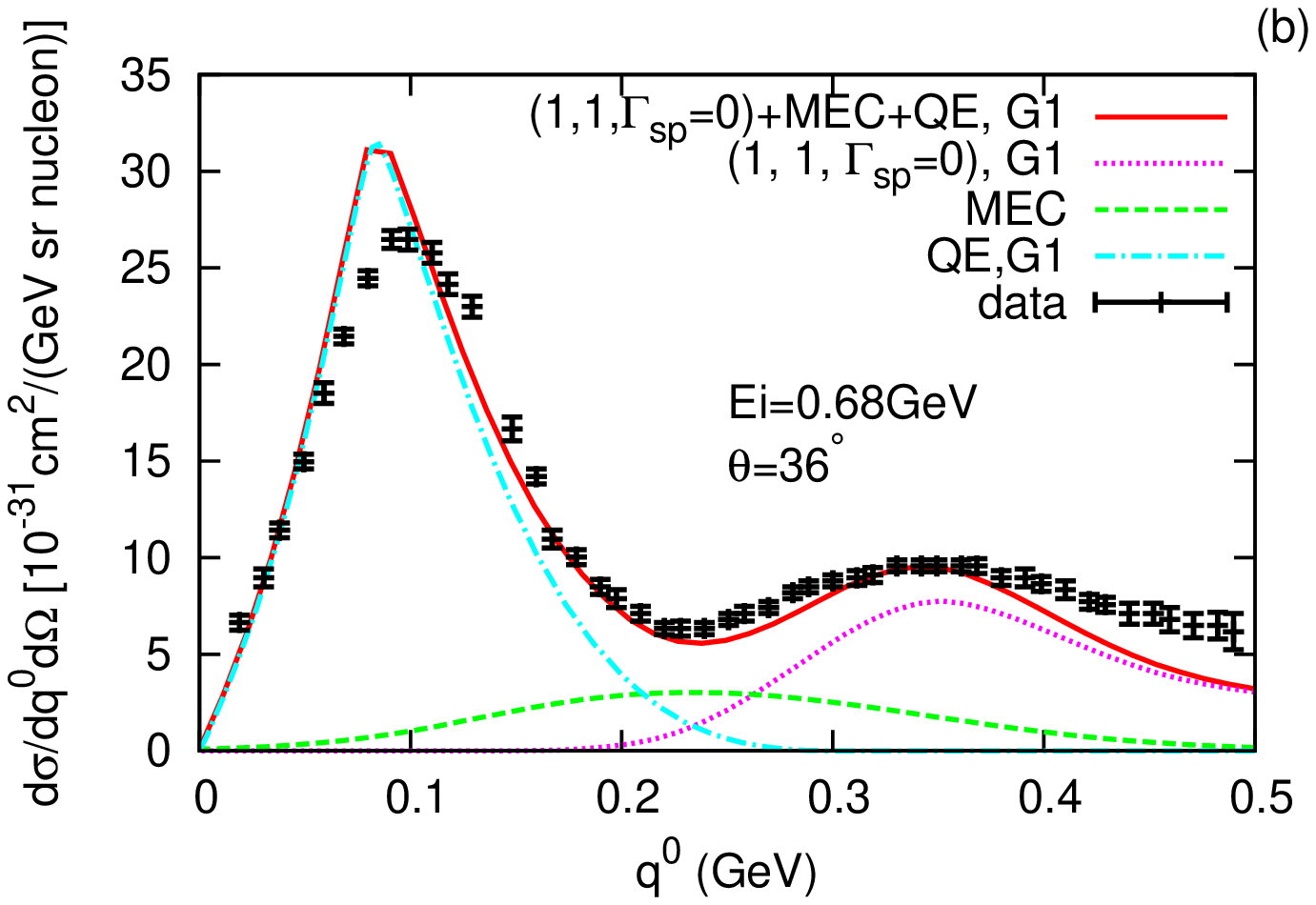}
\caption{(Color online) Inclusive data for differential cross section of 
         electron scattering off ${}^{12}C$. The incoming 
         electron energy is $E_{i}=0.68 \ \mathrm{GeV}$, 
         and the scattering angle is $\theta_{lf}=36^{\circ}$. The kinematics is 
         measured in the laboratory frame.
         The data are from \cite{Barreau83}. Explanations of different plots can be 
         found in the text.
        }
\label{fig:emscattering_0.68_36}
\end{figure}

\begin{figure}
\includegraphics[scale=0.7,angle=-90]{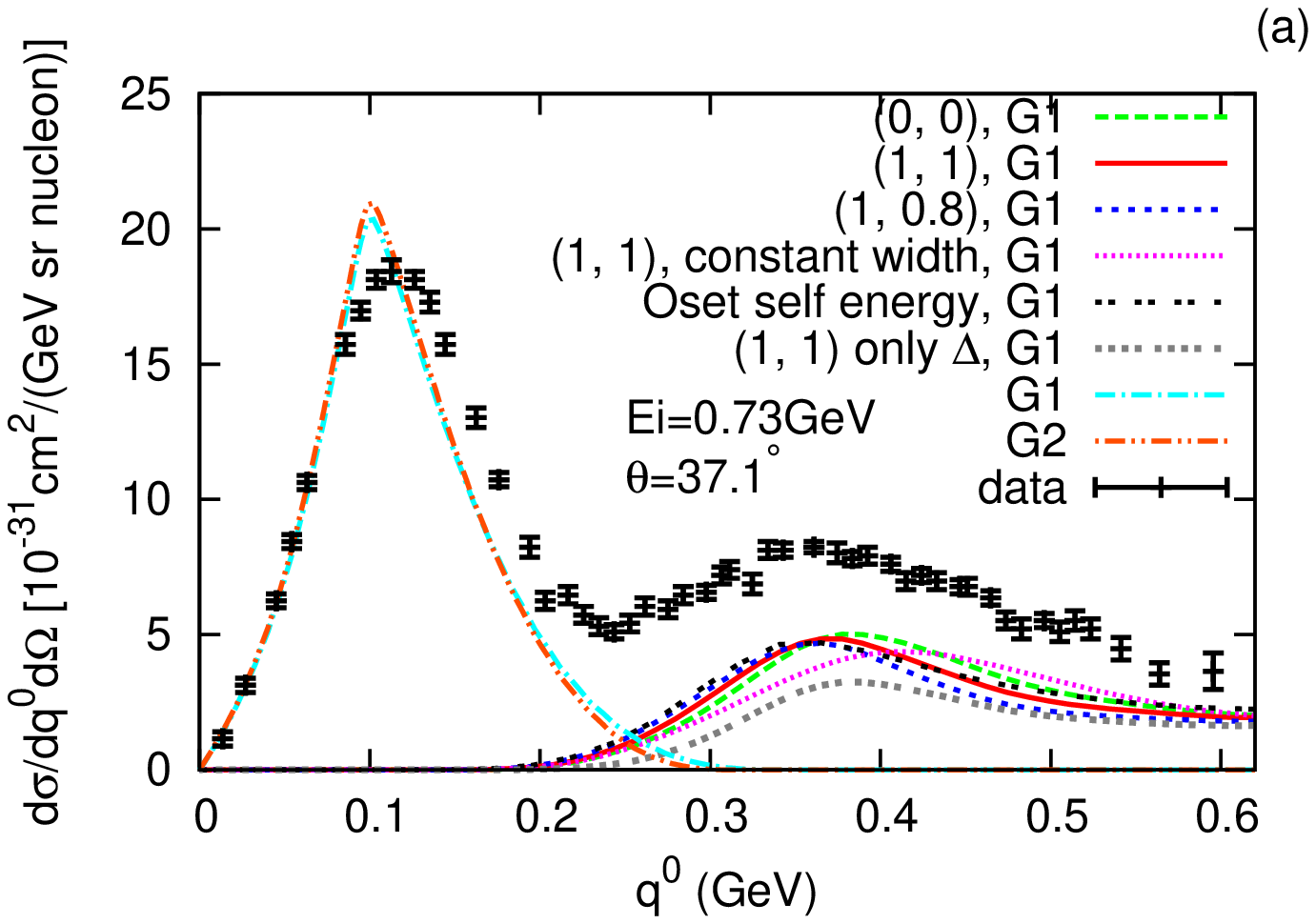}
\includegraphics[scale=0.7,angle=-90]{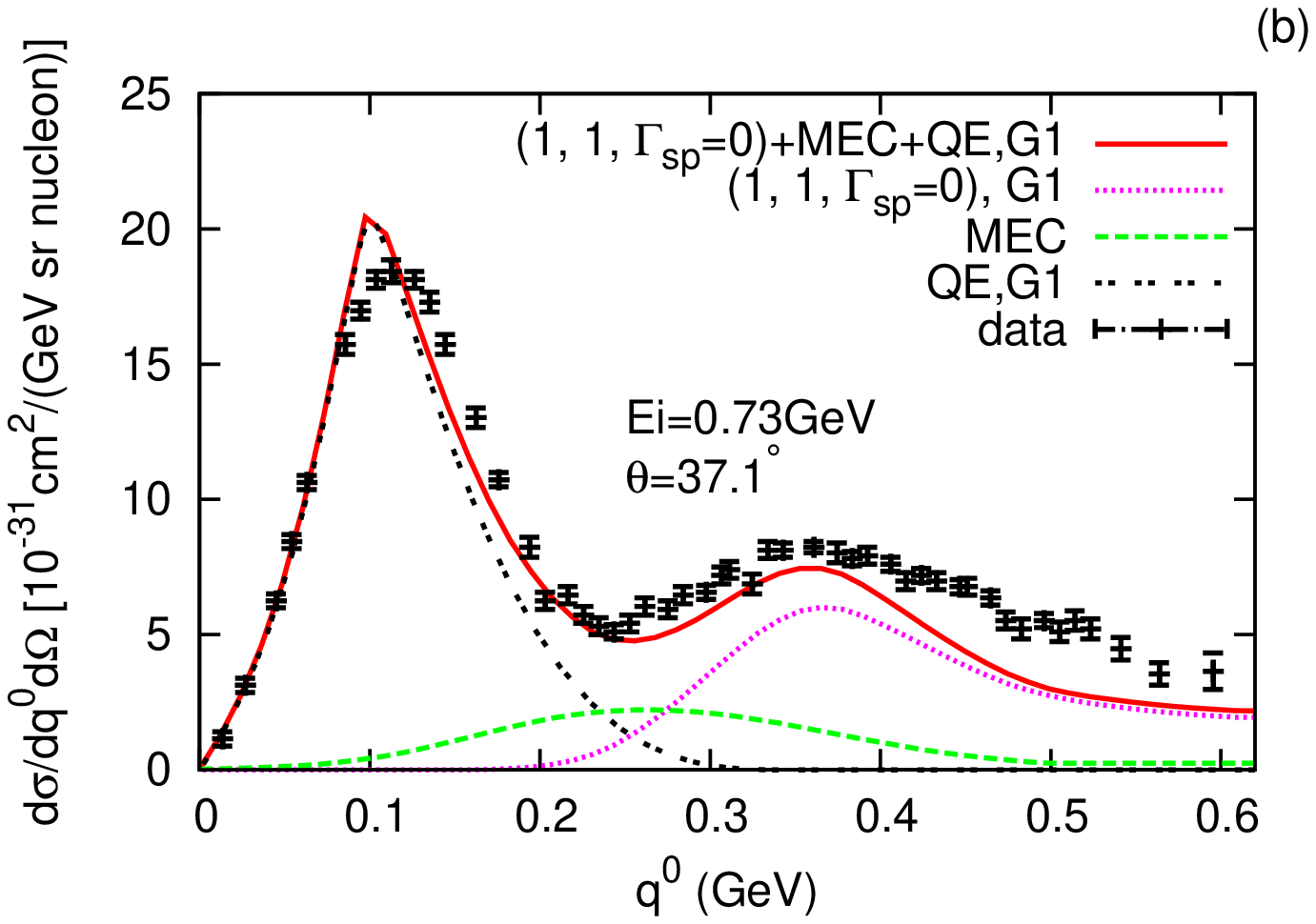}
\caption{(Color online) Inclusive data for differential cross section of 
         electron scattering off ${}^{12}C$. The incoming 
         electron energy is $E_{i}=0.73 \ \mathrm{GeV}$, 
         and the scattering angle is $\theta_{lf}=37.1^{\circ}$. 
         The data are from \cite{Connell87}. Explanations of different plots can be 
         found in the text.
        }
\label{fig:emscattering_0.73_37.1}
\end{figure}

In the upper panels of Figs.~\ref{fig:emscattering_0.62_60}, \ref{fig:emscattering_0.68_36}, and
~\ref{fig:emscattering_0.73_37.1}, we present differential 
cross sections $d\sigma/dq^{0}d\Omega$ for electron scattering 
off ${}^{12}C$ at given electron energies and scattering angles. 
In this section we only focus on the so-called quasi-elastic peak 
at the lower energy region,
which is believed to be dominated by one nucleon knock out. 
The higher energy peak will be discussed in 
Sec.~\ref{sec:pionprod}. 
In Fig.~\ref{fig:emscattering_0.62_60}, the electron energy is 
$E_{i}=0.63 \ \mathrm{GeV}$, and the scattering angle is $\theta_{lf}=60^{\circ}$. 
The plots ``G1'' and ``G2'' are the calculations done with G1 
and G2 parameter sets \cite{Furnstahl9798}. The difference 
between the two is small. The data are from Ref.~\cite{Barreau83}. 
The validity of the form factors realized by 
meson dominance needs to be discussed here. In this figure, 
$Q^{2} \approx 0.3 \ \mathrm{GeV}^{2}$ and 
$\modular{q}{} \approx 0.55 \ \mathrm{GeV}$ at the peak. 
Below the peak, $Q^{2}$ is slightly higher than 
$0.3 \ \mathrm{GeV}^{2}$, and above the peak, 
$Q^{2} \leqslant 0.3 \ \mathrm{GeV}^{2}$. As discussed in \cite{1stpaper}, 
meson dominance
works when $Q^{2} \leqslant 0.3 \ \mathrm{GeV}^{2}$, and hence it can 
be applied here. This is also true for Figs.~\ref{fig:emscattering_0.68_36} 
and \ref{fig:emscattering_0.73_37.1}.
In Fig.~\ref{fig:emscattering_0.68_36}, the electron energy is 
$E_{i}=0.68\ \mathrm{GeV}$, and the scattering angle is $\theta_{lf}=36^{\circ}$. 
In Fig.~\ref{fig:emscattering_0.73_37.1}, the electron energy is 
$E_{i}=0.73 \ \mathrm{GeV}$, and the scattering angle is  $\theta_{lf}=37.1^{\circ}$. 
The data are from \cite{Barreau83} and \cite{Connell87}. 
Again, we see only small differences between the ``G1'' and ``G2'' parameter sets.

\section{pion production} \label{sec:pionprod}

\subsection{Approximation scheme and $\Delta$ dynamics in the nuclear medium} 
\label{subsec:Deltamodification}
\begin{figure}
\includegraphics[scale=0.9]{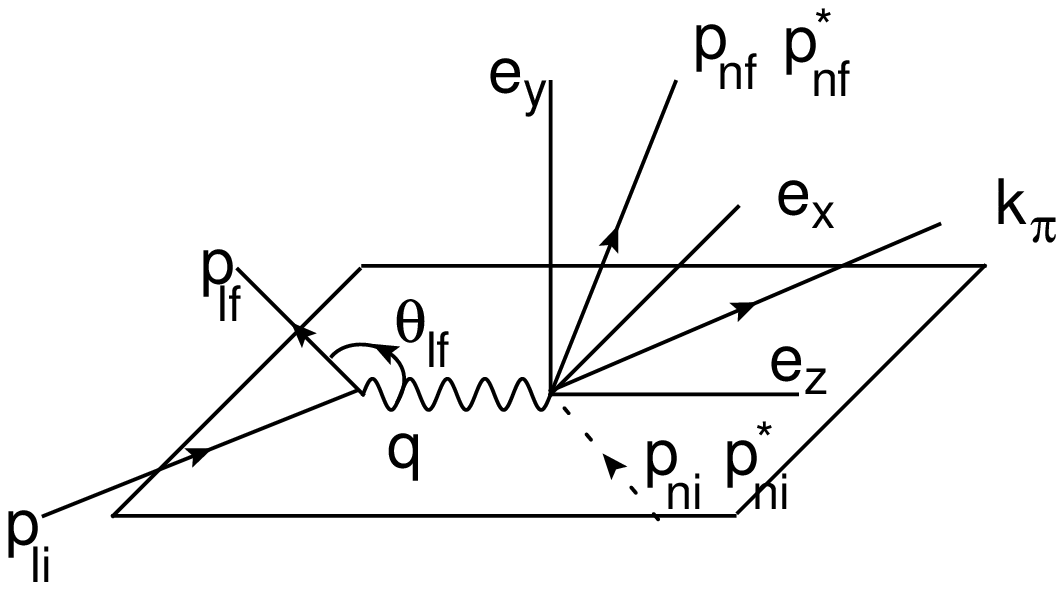}
\caption{The kinematics in the laboratory frame of the nucleus for one nucleon inside the nucleus 
         interacting with the lepton and producing a pion.}
\label{fig:nucleipionprodLabframe}
\end{figure} 
By using the LFG approximation detailed before, the formula for the
cross section can be written as 
\begin{eqnarray}
\sigma &=& \int dV \frac{1}{2 p_{li}^{0}}
           \int \frac{d^{3}{\vec{p}_{nf}}^{\ast}}
                     {(2\pi)^{3} 2 p_{nf}^{\ast 0}} 
                \frac{d^{3}\vec{k}_{\pi}}
                     {(2\pi)^{3} 2 k_{\pi}^{0}}
                \frac{d^{3}\vec{p}_{lf}}{(2\pi)^{3} 2 p_{lf}^{0}} 
                \frac{d^{3}{\vec{p}_{ni}}^{\ast}}
                     {(2\pi)^{3} 2 p_{ni}^{\ast 0}} \notag \\[5pt]
        && \times (2\pi)^{4} 
                  \delta^{4} (q+\starobject{p_{ni}}-\starobject{p_{nf}}-k_{\pi}) 
                  \underset{s_{f},s_{i}}{\sum} 
                  \vert M_{fi} \vert^{2} \ . \label{eqn:xsecpion}
\end{eqnarray}
The details can be found in Appendix~\ref{app:pionprodkinematics}. 
The notations for various momenta are explained in 
Fig.~\ref{fig:nucleipionprodLabframe}. All the integrations 
except the volume integration depend on the space coordinate 
$\vec{r}$ through the space dependent Fermi momentum.
The amplitude $M_{fi}$ in Eq.~(\ref{eqn:xsecpion}) is similar to those 
in Eqs.~(\ref{eqn:matrixelementEMcurrent}), (\ref{eqn:matrixelementCCcurrent}), and (\ref{eqn:matrixelementNCcurrent}), except that
the hadronic currents should be changed to those relevant to pion production: 
\begin{eqnarray}
\langle J^{(had) \mu} \rangle \equiv \bra{N,\pi} J^{(had) \mu} \ket{N}   
\ . \notag 
\end{eqnarray}  
\begin{figure}
\centering
\includegraphics[scale=0.6]{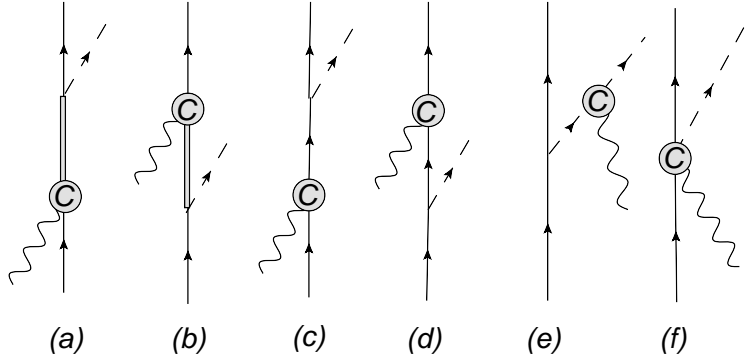}
\caption{Feynman diagrams for
pion production. Here, {\bf C} stands for various types of currents
including vector, axial-vector, and baryon currents. Some diagrams
may be zero for some specific type of current. See Ref.~\cite{1stpaper} for the details.}
\label{fig:feynmanpionproduction}
\end{figure}
Several Feynman diagrams contribute here, as shown in 
Fig.~\ref{fig:feynmanpionproduction}, including diagrams with the $\Delta$ 
[(a) and (b)] and all the rest which we define as nonresonant diagrams. 
See Ref.~\cite{1stpaper} for details about them.  
Among the medium-modifications of the matrix elements, the behavior 
of the $\Delta$ needs to be singled out.    
First, let us focus on the real part of the $\Delta$ self-energy.  
We start from the following Lagrangian (and a similar Lagrangian 
can be found in \cite{wehrgerger93}):
\begin{eqnarray}
\lag_{\Delta; \pi,\rho,V,\phi}
&=&\frac{-i}{2} \psibar{\Delta}^{a}_{\mu} 
                \left\{ \usigma{\mn}\ , \ 
                        \left( i\slashed{\ucpartial{}}
                              -h_{\rho}\slashed{\rho}
                              -h_{v}\slashed{V}
                              -m+h_{s}\phi
                        \right)
               \right\}_{a}^{\mkern3mu b} 
               \Delta_{b\nu} \notag \\ [5pt]
&&{}-\frac{\widetilde{f}_{\rho}h_{\rho}}{4m} 
     \psibar{\Delta}_{\lambda} \rho_{\mn} \usigma{\mn}\Delta^{\lambda}
    -\frac{\widetilde{f}_{v}h_{v}}{4m} 
     \psibar{\Delta}_{\lambda} V_{\mn} \usigma{\mn} \Delta^{\lambda} 
    \ . \label{eqn:deltabg} 
\end{eqnarray}
Here the $\Delta$ field is given by the Rarita-Schwinger representation, and 
$a, b =\pm 3/2,\pm 1/2 $ are $\Delta$ isospin indices \cite{bookchapter}. 
At the normal nuclear density, the $\Delta$ is not stable in the nuclear medium. So 
the expectation values of meson fields are not changed in normal nuclei at the  mean-field level. 
Similar to the nucleon case, the $\Delta$ spectrum in nuclear matter (without the $\Delta$-pion interaction) is given by
\begin{eqnarray}
p_{\Delta}^{0} &=& h_{v} \langle V^{0} \rangle 
                  +\sqrt{{\mstar}^{2}+\vec{p}_{\Delta}^{2}} \notag \\[5pt]
          &\equiv& h_{v} \langle V^{0} \rangle + p_{\Delta}^{\ast 0} 
          \notag \\[5pt]
               &=& h_{v} \langle V^{0} \rangle 
                  +\sqrt{{\mstar}^{2}+\vec{p}_{\Delta}^{\ast 2}} 
              \ ,  \notag \\[5pt]
\mstar  & \equiv & m-h_{s} \langle \phi \rangle \ . \notag 
\end{eqnarray}
The effect of introducing $h_{s}$ and $h_{v}$ couplings 
on the equation of state (EOS) was analyzed in 
\cite{wehrgerger89,boguta81,kosov98}. Some constraints 
on the couplings, $r_{s}\equiv h_{s}/g_{s}$ and 
$r_{v}\equiv h_{v}/g_{v}$, were calculated in 
\cite{wehrgerger89,kosov98}. Here we resort to 
the scattering problem to find other constraints. In pion-nucleus scattering studies 
\cite{horikawa80,Nakamura10}, S-L coupling of the $\Delta$ inside the 
nucleus was introduced by hand, although 
its origin is not clear in the nonrelativistic model. 
In this model, a mechanism similar to 
the generation of the nucleon's S-L coupling is used to generate $\Delta$'s. 
Following discussions in \cite{Furnsthal98} and using the 
Lagrangian in Eq.~(\ref{eqn:deltabg}), we can estimate the 
S-L coupling of the $\Delta$ as
\begin{eqnarray}
h_{\Delta} &=& \frac{1}{3} \left[ \frac{1}{2\psibar{m}^{2} \ r} 
                                  \frac{d}{dr} 
                                  \left( h_{s}\langle\phi\rangle
                                        +h_{v}\langle V^{0}\rangle\right)
                                        -\frac{\widetilde{f}_{v}}
                                              {m\psibar{m} \ r}
                                         \frac{d}{dr} 
                                         \left(h_{v}\langle V^{0} \rangle\right)
                          \right] \vec{S} \cdot \vec{L}  
                          \notag \\[5pt]
       &\equiv& \alpha(r)  \vec{S} \cdot \vec{L} \ . \label{eqn:lscoupling} 
\end{eqnarray}
Here, $\psibar{m}\equiv m-\half \left(h_{s}\langle\phi\rangle+h_{v}\langle 
V^{0} \rangle\right)$. 
\begin{figure}
\includegraphics[scale=0.7, angle=-90]{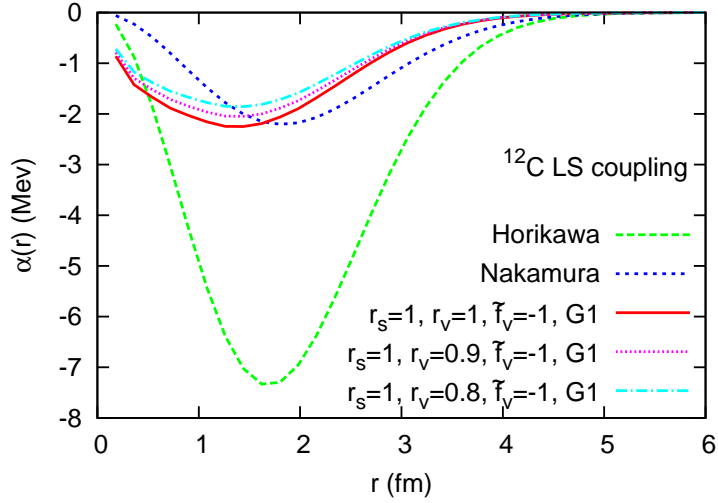}
\caption{(Color online) The strength of $\Delta$'s S-L coupling in ${}^{12}C$. 
          Here we compare two phenomenology 
          results with our three calculations based on different 
          parameter sets. The ``Horikawa'' is from \cite{horikawa80}. 
          The ``Nakamura'' is from \cite{Nakamura10}. All these 
          calculations involve setting the G1 parameter to describe 
          the nucleus ground state. We change $r_{v}$ to $1, \ 0.9, \ 0.8$ 
          while keeping $r_{s}=1, \ \widetilde{f}_{v}=-1$.
        }
\label{fig:LS_c12} 
\end{figure}
In Fig.~\ref{fig:LS_c12}, we compare our estimates of $\alpha(r)$ 
defined in Eq.~(\ref{eqn:lscoupling}) with two different phenomenological 
fits for ${}^{12}C$. We can see that our 
estimates based on three different parameter sets, 
$r_{s}=1, \ r_{v}=1, \ 0.9, \ 0.8, \ \text{and} \ \widetilde{f}_{v}=-1.0$, 
are consistent with the ``Nakamura'' result in \cite{Nakamura10}, while the 
``Horikawa'' result in \cite{horikawa80} is significantly larger than the 
``Nakamura'' result when $r\geq 1 \ \mathrm{fm}$. 
Meanwhile, all the couplings are consistent with the ``naturalness'' 
assumption, which also motivates our choice of 
$\widetilde{f}_{v}=-1$. We do not show the consequence of $r_{s}=r_{v}=0$, 
since there is no S-L coupling generated in this case.

Second, we turn to the imaginary part of the self-energy. It is known 
that Pauli blocking effects decrease the
width due to reduction of the pion-decay phase space, 
while the collision channels, $\Delta N \leftrightarrow N N$ for example, 
increase the width. 
The two competing processes have been 
investigated in nonrelativistic models.
At normal nuclear density, the 
net result is to increase the width \cite{oset87}. 
In phenomenological fits \cite{hirata79} \cite{horikawa80}, 
this increase is taken into account by introducing a density-dependent 
complex spreading potential for the $\Delta$. 
Here we follow this approach. Above the pion threshold, 
\begin{eqnarray}
\Gamma_{\Delta}&=&\Gamma_{\pi} + \Gamma_{\mathrm{sp}}  \ , \notag \\
\Gamma_{\mathrm{sp}}&=&V_{0}\times \frac{\rho(r)}{\rho(0)} \ .
\label{eqn:phenomenology_Delta_width1}  
\end{eqnarray}
$\Gamma_{\pi}$ is the $\Delta$ pion-decay width \cite{oset87, wehrgerger90}.
\footnote{In the $\Gamma_{\pi}$ calculation, only Pauli-blocking is considered. 
Modifications of the real part of the nucleon and $\Delta$ self-energies are not included.}
$\Gamma_{\mathrm{sp}}$ ($V_{0} \approx 80 \ \mathrm{MeV}$) 
is the width in other channels, and it 
has been fitted in \cite{horikawa80, Nakamura10}. 
Below the pion threshold (which is useful in photon production), 
\begin{eqnarray}
\Gamma_{\Delta}=\Gamma_{\mathrm{sp}}=V_{0}\times \frac{\rho(r)}{\rho(0)} \ . 
\label{eqn:phenomenology_Delta_width2}
\end{eqnarray}
In the cross channel of the $\Delta$ diagram, we set the width to zero. 
Moreover, in the literature \cite{Praet09,MacGregor98}, the simple increase of the 
$\Delta$ width by $\delta \Gamma \approx 40 \ \mathrm{MeV} $ has been used
for pion production:
\begin{eqnarray}
\Gamma_{\Delta} &\rightarrow&                     
120  +40 \ \mathrm{MeV} \ . \label{eqn:phenomenology_Delta_width3} 
\end{eqnarray}
This procedure turns out to work qualitatively, as will be shown later.
Furthermore, in \cite{Gil97}, 
the $\Delta$ self-energy calculated in \cite{oset87} is used for 
inclusive electron scattering off nuclei. 
In Sec.~\ref{subsection:eletroproduction}, 
we will compare our results using 
Eqs.~(\ref{eqn:phenomenology_Delta_width1}) 
and (\ref{eqn:phenomenology_Delta_width2}) with those using Eq.~(\ref{eqn:phenomenology_Delta_width3}) and the width in \cite{Gil97, oset87}.

\subsection{Pion electroproduction} \label{subsection:eletroproduction}

Here we focus on the region beyond quasi-elastic 
scattering in the upper panels of Figs.~\ref{fig:emscattering_0.62_60}, \ref{fig:emscattering_0.68_36}, 
and~\ref{fig:emscattering_0.73_37.1}. It is 
believed that the second peak mainly comes from the $\Delta$ 
excitation inside the nucleus. In the upper panels of these figures, 
we provide our pion-production 
 results (without FSI) due to six different calculations.  
We include the full set of Feynman diagrams in the first five calculations, 
and diagrams with the $\Delta$ in $s$ and $u$ channels in the sixth. The difference among 
the first three calculations is the choice of $(r_{s}, \ r_{v})$
parameter sets: $(r_{s}=0, \ r_{v}=0)$, $(r_{s}=1, \ r_{v}=1)$, 
and $(r_{s}=1, \ r_{v}=0.8)$. In these three, the $\Delta$ 
width shown in Eqs.~(\ref{eqn:phenomenology_Delta_width1}) and 
(\ref{eqn:phenomenology_Delta_width2}) is applied. In the fourth calculation, 
we set $(r_{s}=1, \ r_{v}=1)$ and apply the constant shift of the $\Delta$ width 
as shown in Eq.~(\ref{eqn:phenomenology_Delta_width3}). The fifth calculation is 
done by using the $\Delta$ self-energy as 
calculated in \cite{oset87, Nieves93}, which is essentially repeating the calculations 
in \cite{Gil97}. The sixth calculation has $(r_{s}=1, \ r_{v}=1)$ and uses the 
same $\Delta$ width as used in the first three.  

First, let us discuss the location of the $\Delta$-peak along the $q^{0}$ axis. 
The different choice of $r_{s}$ and $r_{v}$ indicates different binding 
potentials for $\Delta$. For $(0, \  0)$, the real part of the self-energy is the same as in vacuum without any binding. For $r_{s}=1$, $\Delta$ has the same attractive potential 
as the nucleon. The vector part tuned by $r_{v}$ provides a 
repulsive potential. So we can see that $(1, \  0.8)$ has a deeper binding 
potential than $(1, \ 1)$. Hence, $(0, \  0)$ is less bound than $(1, \  1)$ and $(1, \  1)$ is less bound than $(1, \  0.8)$. 
In Figs.~\ref{fig:emscattering_0.62_60}, \ref{fig:emscattering_0.68_36}, 
and \ref{fig:emscattering_0.73_37.1}, the location of the $\Delta$ peak in first three 
calculations indeed follows this argument. 
(We can estimate the location of the $\Delta$ peak in a global Fermi gas model. 
\footnote{Following \cite{{Rosenfelder80}}, for ${}^{12}C$, we assume a 
global Fermi gas, with the nucleon effective mass $\Mstar=0.75 M$, 
$g_{s}\langle\phi\rangle \approx0.235 \ \mathrm{GeV}$, 
$g_{v}\langle V^{0}\rangle\approx 0.75 g_{s}\langle\phi\rangle$ (in the laboratory frame), and 
$\mstar=m-r_{s}g_{s}\langle\phi\rangle\approx0.995 \ \mathrm{GeV} (r_{s}=1)$. 
Meanwhile in the $s$ channel, the $\Delta$ momentum is 
$p^{\ast}_{\Delta}=q+p^{\ast}_{ni}+(1-r_{v})g_{v}\langle V^{0}\rangle$. 
It is easy to calculate the $q^{0}$-location of the peak by setting 
$p^{\ast 2}_{\Delta}=m^{\ast2}$. For $E_{i}=0.62 \ \mathrm{GeV}$ and 
$\theta=60^{\circ}$ (see Fig.~\ref{fig:emscattering_0.62_60}), 
$q^{0}=0.43 \ \mathrm{GeV}$ if $r_{v}=1$ 
and $q^{0}=0.39 \ \mathrm{GeV}$ if $r_{v}=0.8$.}) 
The fourth calculation with $(1, \  1)$ and constant width does not give the 
correct peak position in the three figures: It underestimates (overestimates) $\Delta$'s 
contribution on the left (right) side of the peak, because the constant width 
assumption overestimates (underestimates) $\Delta$'s width on the left (right) side. 
The fifth calculation by using $\Delta$ modification calculated  
in \cite{oset87, Nieves93} gives the correct location of the peak. 
Comparing the second with the sixth calculation, we can see the 
significance of nonresonant contributions (they use the same set of parameters and the $\Delta$ width). 

However, the pion production channel could not explain the full 
strength of the $\Delta$ peak. Meanwhile in the
``dip'' region between the quasi-elastic scattering and the  
$\Delta$ peak, the calculations also miss strength. 
This indicates we miss other channels from dip region to the
$\Delta$ peak. Missing strength at the \emph{peak} position 
can be qualitatively explained by considering the fourth 
calculation in the upper panels, 
whose simple treatment of the $\Delta$ width makes analysis 
transparent. According to Eqs.~(\ref{eqn:phenomenology_Delta_width1}) and 
(\ref{eqn:phenomenology_Delta_width3}), we estimate $0.04  \leqslant \Gamma_{\mathrm{sp}} 
\leqslant 0.08 \ \mathrm{GeV}$ in the sense of averaging over ${}^{12}C$, 
and hence $0.08 \leqslant\Gamma_{\pi} \leqslant 0.12 \ \mathrm{GeV}$. 
The comparable width of other decay channels shows their importance to the inclusive data.
Moreover, there are contributions from two-body currents without $\Delta$ as an intermediate state. 
In the lower panels of Figs.~\ref{fig:emscattering_0.62_60}, \ref{fig:emscattering_0.68_36}, 
and~\ref{fig:emscattering_0.73_37.1}, we add up three different channels: quasi-elastic, 
pion production, and two-body current contributions 
[labeled as meson-exchange-current (MEC) in the plots]. 
The label ``$(1, \ 1, \ \Gamma_{\mathrm{sp}}=0)$'' 
for pion production applies under the
assumption that $(r_{s}=1, \ r_{v}=1)$ and that no new channel takes away the flux 
from the $\Delta$ pion production. The MEC-contributions are 
from \cite{DePace03} \cite{donnelly2012}. Here the total strength matches well 
with the inclusive data. 
However a detailed study of different channels in the QHD EFT framework 
is needed to address this issue conclusively.  

The difference between our calculations and those in \cite{wehrgerger89} where the QHD model is also applied should be mentioned here. 
Ours are strictly 
based on the field theory, while in \cite{wehrgerger89} the $\Delta$ is introduced by hand (where they were convoluted with the cross section 
based on a ``stable'' $\Delta$ theory with a Lorentzian weight 
function). Moreover, we take into account the contribution from 
other diagrams, which are not considered in \cite{wehrgerger89}. 
The two results are different somewhat, but our choice
$r_{s}=1, r_{v}=1, \ 0.8$ is consistent with the analysis in 
\cite{wehrgerger89}. 

Moreover, we can see that the differences in cross sections obtained  
using $r_{s}=1, r_{v}=1, \ 0.8$ are not significant, 
which indicates that the total cross sections of neutrinoproduction processes 
are not sensitive to them either.
This will be confirmed by the results in Sec.~\ref{subsec:CCNCpion}.

\subsection{CC and NC pion production} \label{subsec:CCNCpion}

\begin{figure}
\centering
\includegraphics[scale=0.66, angle=-90]{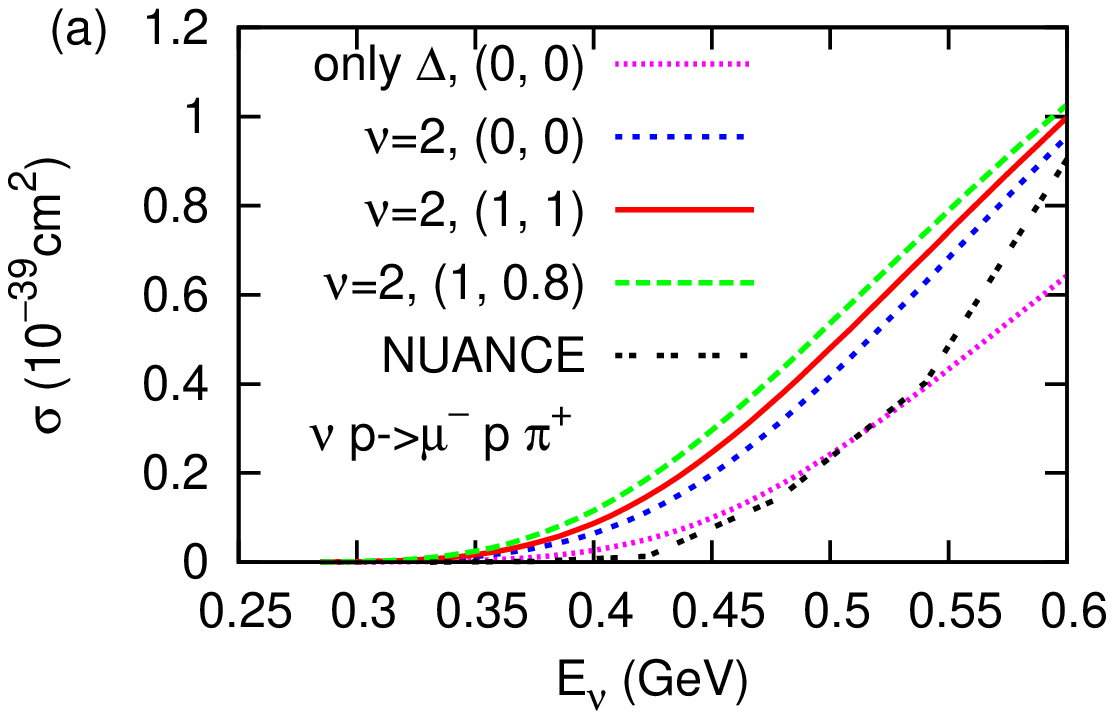}
\includegraphics[scale=0.66, angle=-90]{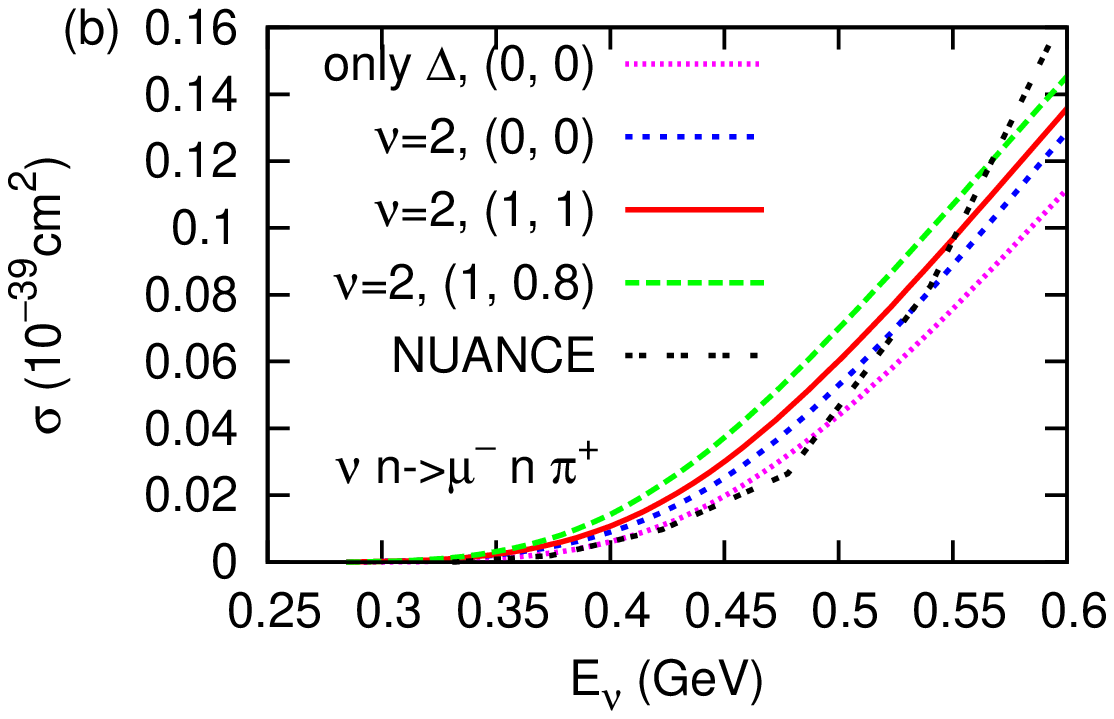}
\includegraphics[scale=0.66, angle=-90]{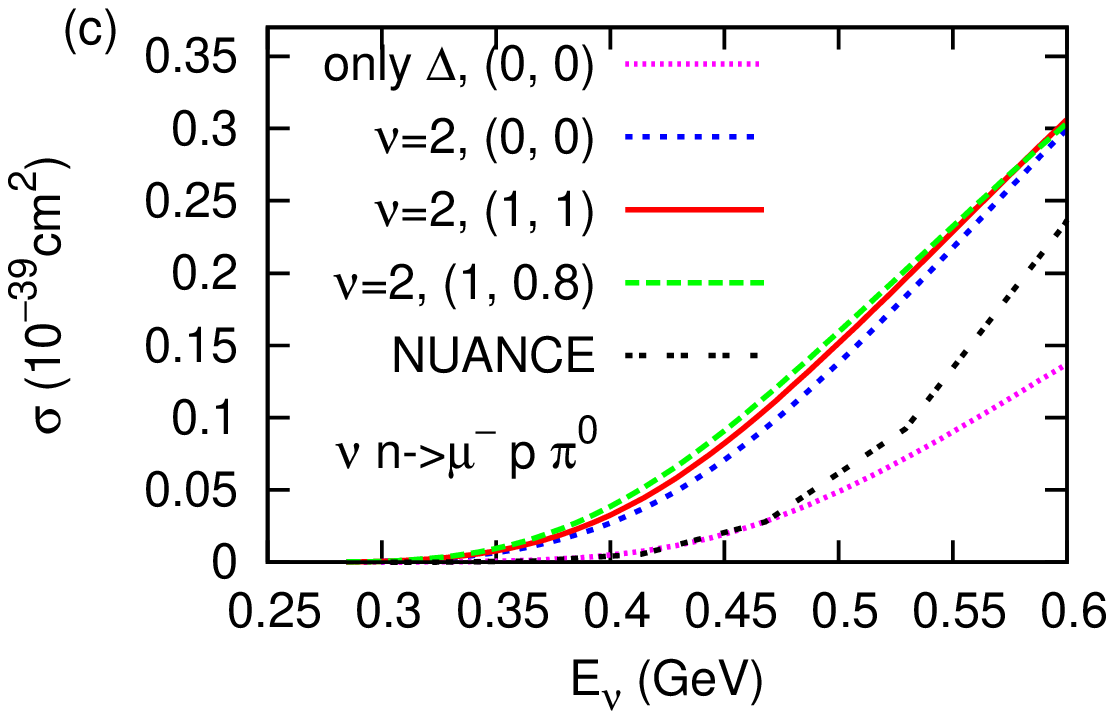}
\includegraphics[scale=0.66, angle=-90]{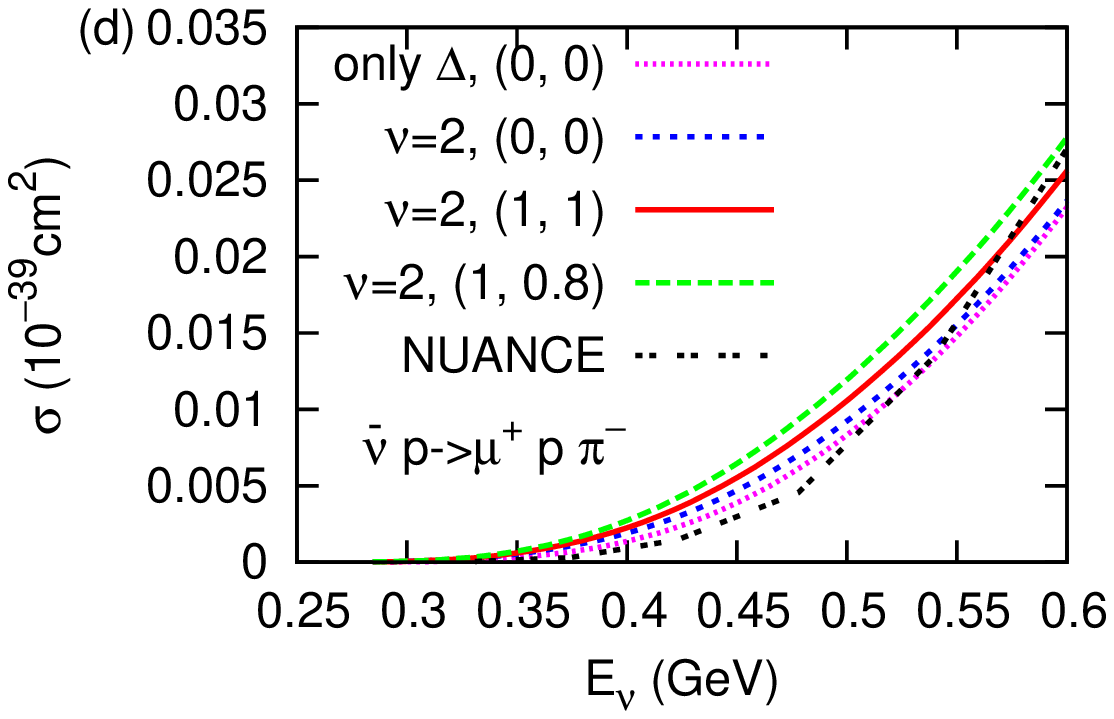}
\includegraphics[scale=0.66, angle=-90]{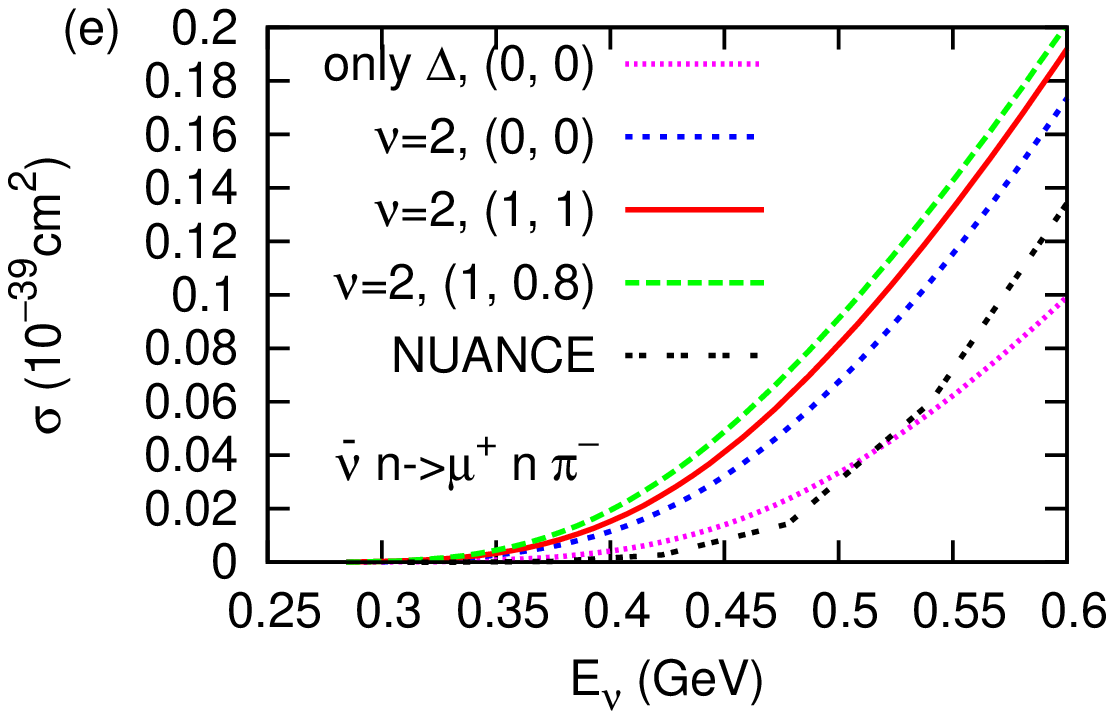}
\includegraphics[scale=0.66, angle=-90]{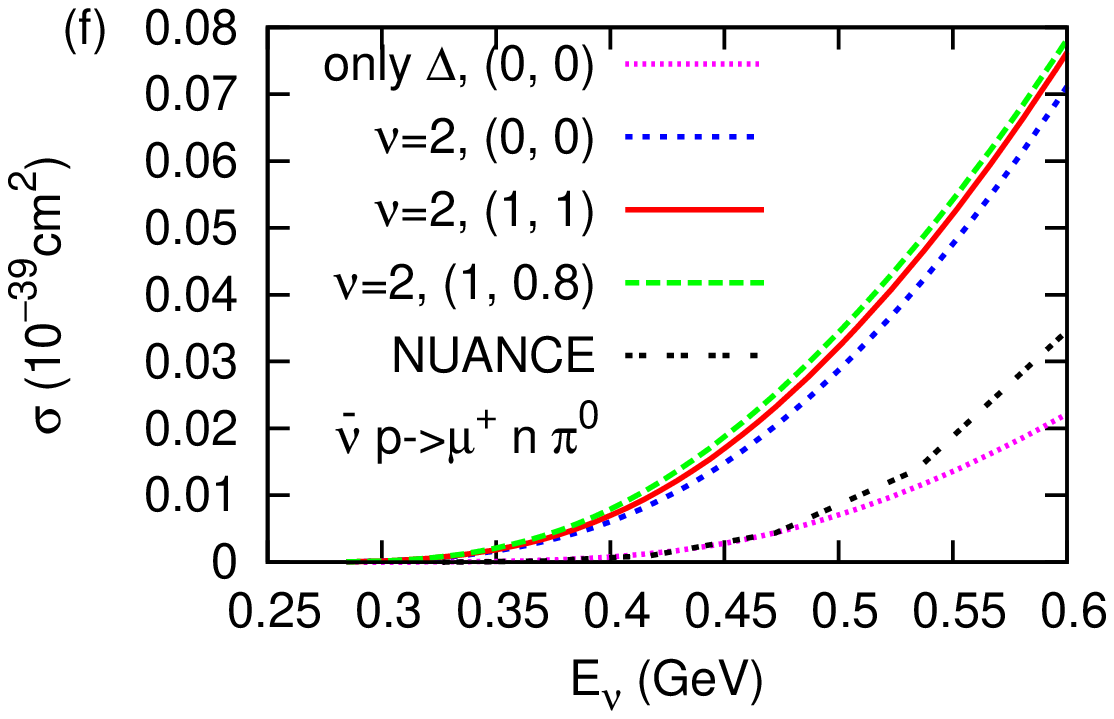}
\caption{(Color online) Total cross section per proton or neutron for the incoherent 
 CC pion production in neutrino-- and antineutrino--${}^{12}C$ scattering.
        }
\label{fig:p_ppion+_c12}
\end{figure}  

Fig.~\ref{fig:p_ppion+_c12} shows the total cross section averaged 
over proton or neutron number for CC pion production 
in (anti)neutrino--${}^{12}C$ scattering. 
We also compare our result with NUANCE's output without FSI. 
In each figure, our calculations  
including different diagrams and using different $r_{s}$ and $r_{v}$
are shown. 
The ``only $\Delta$'' calculation only takes into account $\Delta$ diagrams. 
In the others, all the diagrams up to  $\nu=2$ are included. 
 Systematically in all the channels, 
our ``only $\Delta$'' calculation is close to the NUANCE output.
But other diagrams contained in ``$\nu=2$'' calculations 
are not negligible in all the channels around the resonance region,
especially when the $s$-channel $\Delta$ contribution is 
suppressed by the small Clebsch-Gordan coefficients 
(for example, $\nu_{\mu}+n\longrightarrow \mu^{-}+p+\pi^{0}$). 
In the very low energy region away from the resonance, nonresonant 
diagrams dominate. See \cite{1stpaper} for the power counting of diagrams. 
Moreover, we check that the contributions of 
higher order ($\nu \geqslant 3$) diagrams are tiny. 
We also can see that below 0.5 GeV, the 
$(r_{s}=1 \ , r_{v}=0.8)$ results are bigger than the $(1, \  1)$ results and 
the $(1, \  1)$ results are bigger than the $(0, \  0)$ results. 
Following the discussions in Sec.~\ref{subsection:eletroproduction} 
about the location of the $\Delta$ peak in pion electroproduction, 
we expect that, at a given energy, $\Delta$ excitation occupies 
more phase space in  $(1, \  0.8)$ than in $(1, \  1)$ and more in 
$(1, \  1)$ than in $(0, \  0)$. So the pattern among the three different 
calculations is consistent with the qualitative analysis. Here $(0, \  0)$ 
is presented simply for the purpose of comparison, and its conflict with the $\Delta$
S-L coupling is presented in Sec.~\ref{subsec:Deltamodification}.  

One question needs to be raised: Do we have $\Delta$-dominance 
in the \emph{nuclear} scattering around $0.5$ GeV ? 
If we compare the ``$\nu=2$'' calculations 
with ``only $\Delta$'' calculations in every channel, 
the answer is no. It turns out that the $\Delta$ contribution is strongly 
reduced due to the broadening of its width, 
compared to its contribution in free nucleon scattering. 
Meanwhile nonresonant contributions 
are reduced by Pauli blocking. To see this qualitatively, 
compare our results here with the cross sections shown in 
\cite{1stpaper} for production from free nucleons. 
In \cite{1stpaper}, two different calculations can be found,
including ``Only $\Delta$'' and ``$\nu=2$''. 
We just show the total cross sections at $E_{\nu}=0.5$ GeV in 
Tab.~\ref{tab:freeboundCC} for neutrino scattering. 
For example, ``$p,p\pi^{+}$'' indicates the channel 
$\nu+p\rightarrow \mu^{-}+p+\pi^{+}$. ``(f)'' and ``(b)'' correspond 
to scattering from free nucleons and from bound nucleons 
in ${}^{12}C$ respectively. In both ``only $\Delta$ (b)'' and ``$\nu=2$ (b),'' 
$r_{s}=r_{v}=1$ (and note that calculations with only $\Delta$ and $r_{s}=r_{v}=1$ are not 
shown in the figures).
``Nonresonant (b)'' is the difference between the two, and can be viewed 
qualitatively as the contributions of the nonresonant diagrams. \footnote{%
In principle, there are interferences between contributions from $\Delta$ and 
other diagrams. At $E_{\nu}=0.5$ GeV, we can assume in most of 
phase space that the $\Delta$ is ``on shell'' while contributions from other diagrams 
are real, and hence the interferences are small.} 
The labeling for free nucleon scattering is the same. 
We can see that the $\Delta$ contribution in nuclear scattering 
has been reduced systematically by around 50\% in all channels, 
compared to its contribution in nucleon scattering;
the nonresonant contributions are also strongly reduced. 
Clearly, the nonresonant contributions are not negligible in both 
nucleon and nuclei scattering. 
The same situation occurs in the antineutrino scattering channels and hence 
are not shown explicitly. This underscores the importance of including
nonresonant contributions in CC pion production.
\begin{table}
 \centering
   \begin{tabular}{|c|c|c|c|c|c|c|} \hline
 $\sigma(10^{-39}\mathrm{cm}^{2})$        & only $\Delta$ (f)
           & $\nu=2$ (f)
           & Nonresonant (f)
           & Only $\Delta$ (b)
           & $\nu=2$ (b)
           & Nonresonant (b)   \\  \hline
$p,p\pi^{+}$ & $0.56$
           & $0.85$
           & $0.29$
           & $0.33$
           & $0.48$
           & $0.15$  \\   \hline
$n,n\pi^{+}$ & $0.088$
           & $0.105$
           & $0.017$
           & $0.056$
           & $0.060$
           & $0.004$  \\   \hline   
$n,p\pi^{0}$ & $0.117$
           & $0.258$
           & $0.141$
           & $0.069$
           & $0.153$
           & $0.084$  \\   \hline                     
   \end{tabular}
   \caption{Total cross sections averaged over number of protons or nucleons for CC pion production in neutrino--${}^{12}C$ scattering at $E_{\nu}=0.5$ GeV. See the text for detailed explanations. In the nuclear scattering,
   $r_{s}=r_{v}=1$.} \label{tab:freeboundCC}
\end{table}

\begin{figure}
\begin{center}
\includegraphics[scale=0.66, angle=-90]{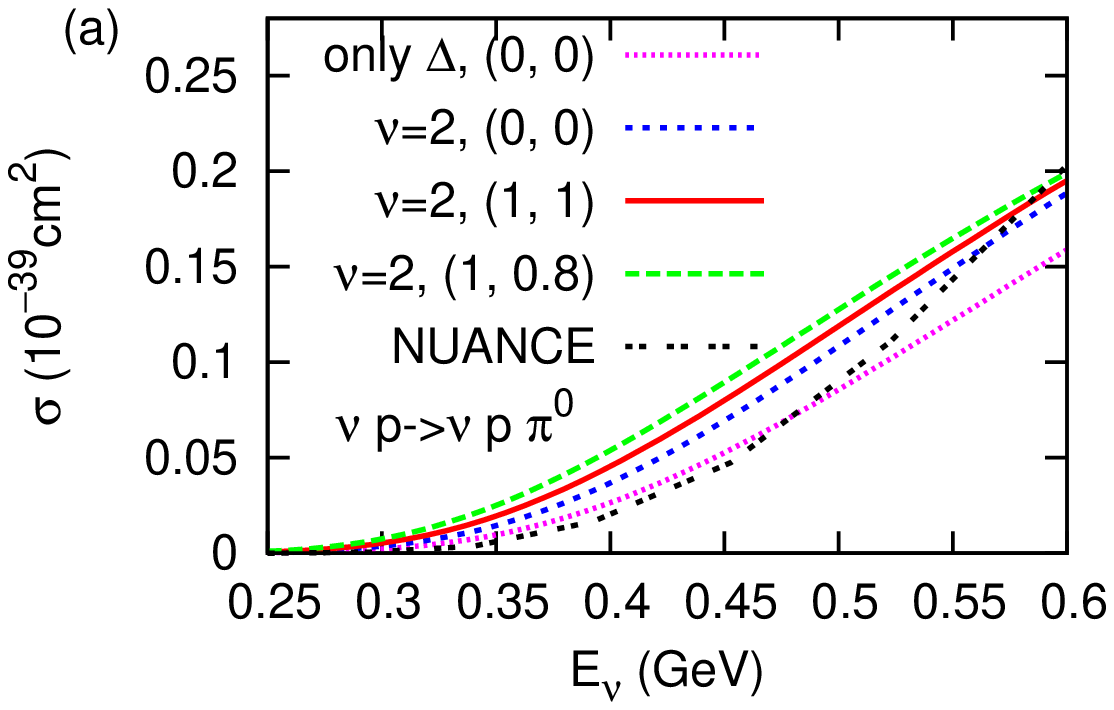}
\includegraphics[scale=0.66, angle=-90]{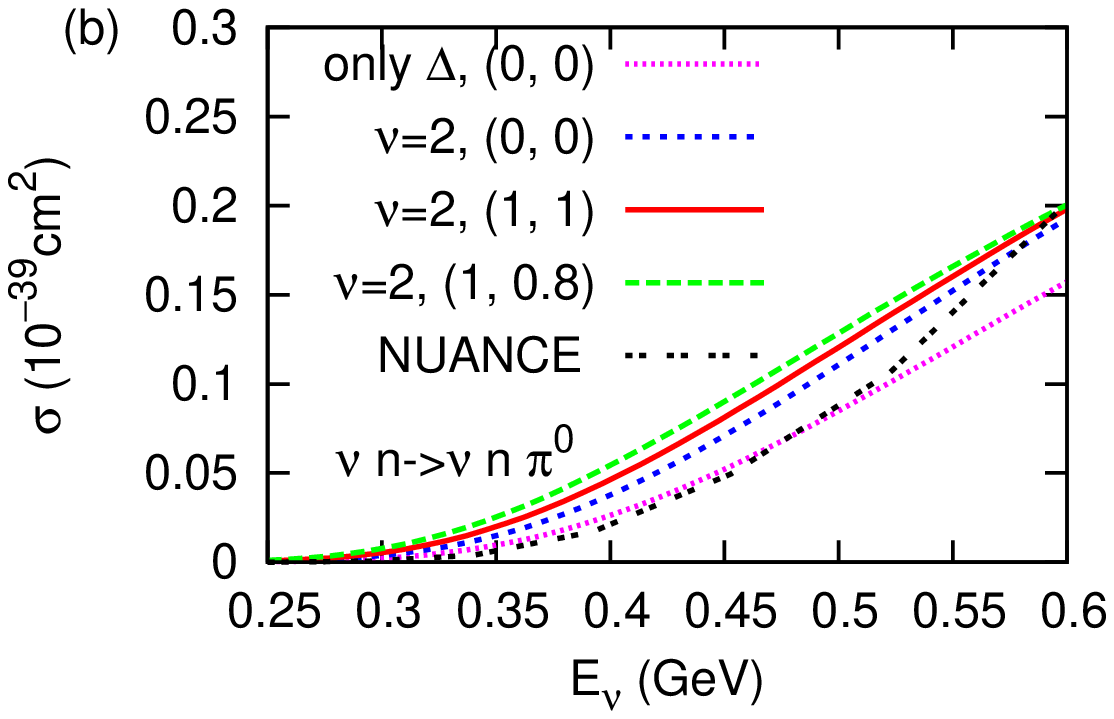}
\includegraphics[scale=0.66, angle=-90]{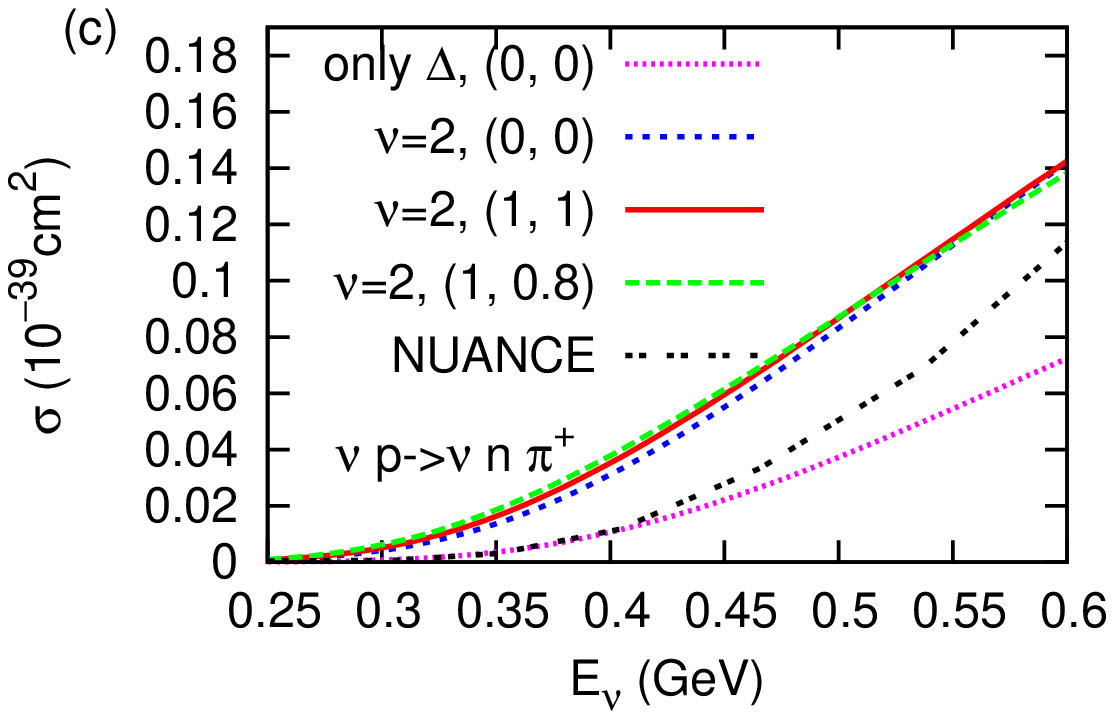}
\includegraphics[scale=0.66, angle=-90]{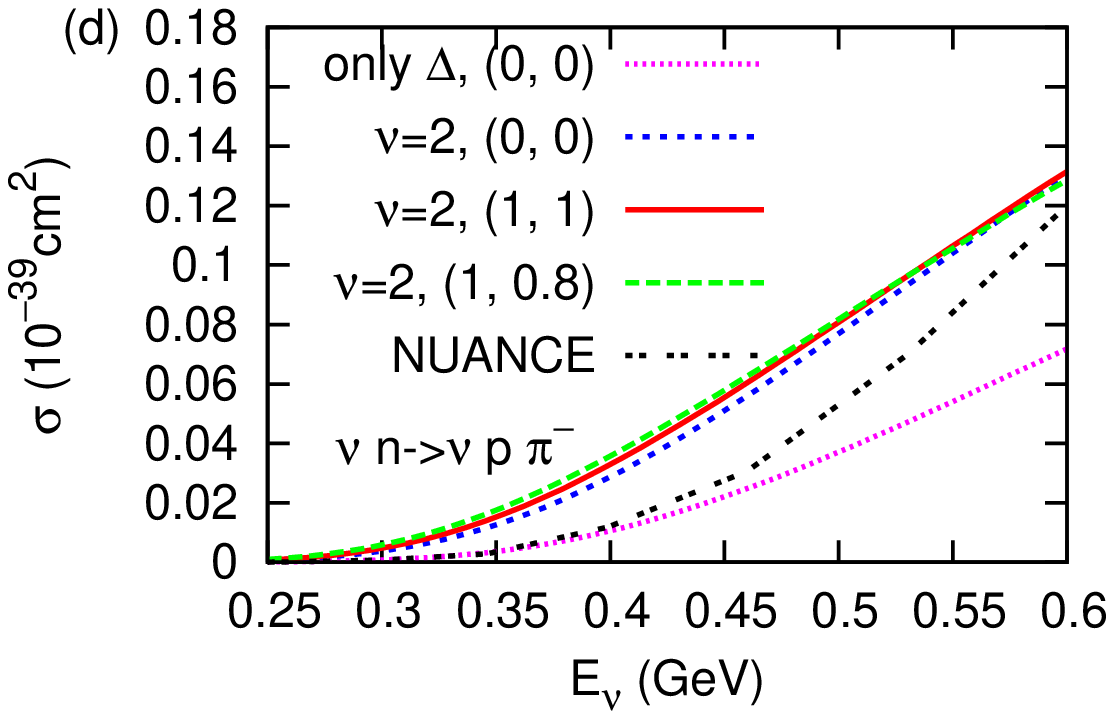}
\includegraphics[scale=0.66,angle=-90]{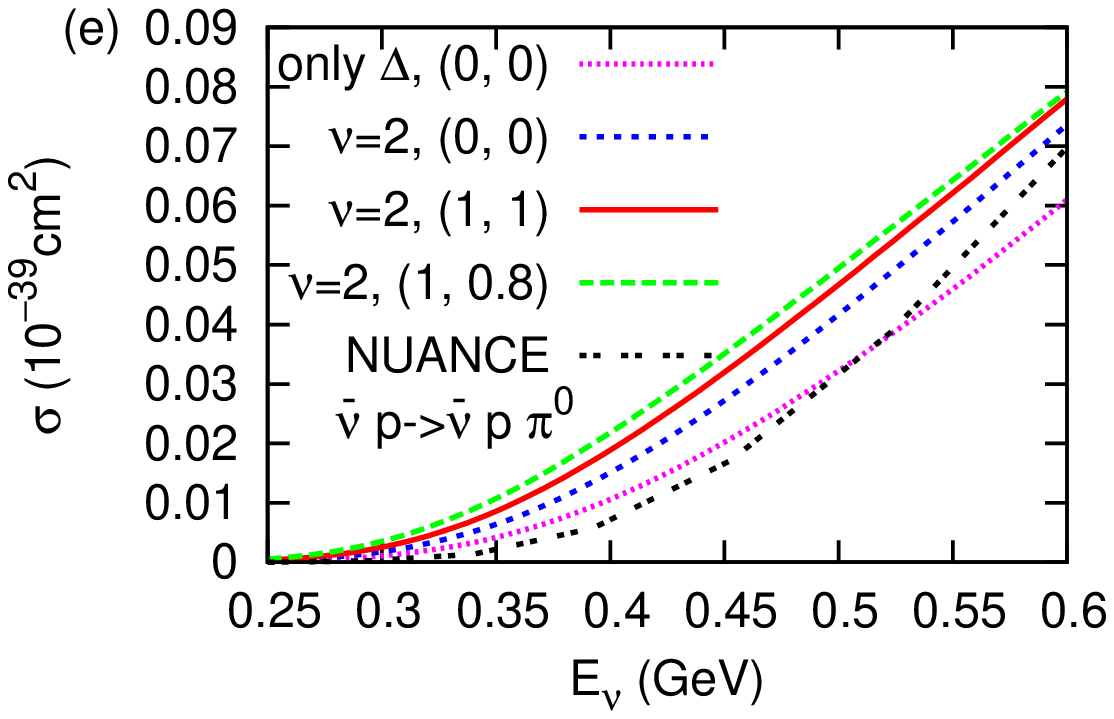}
\includegraphics[scale=0.66,angle=-90]{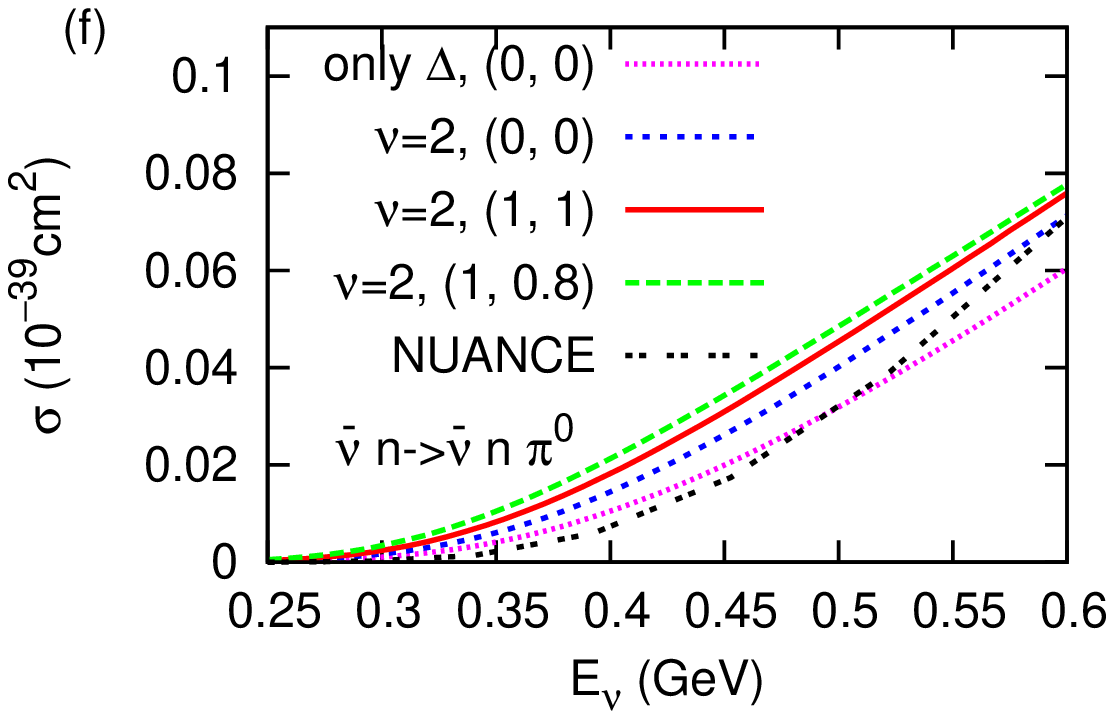}
\includegraphics[scale=0.66,angle=-90]{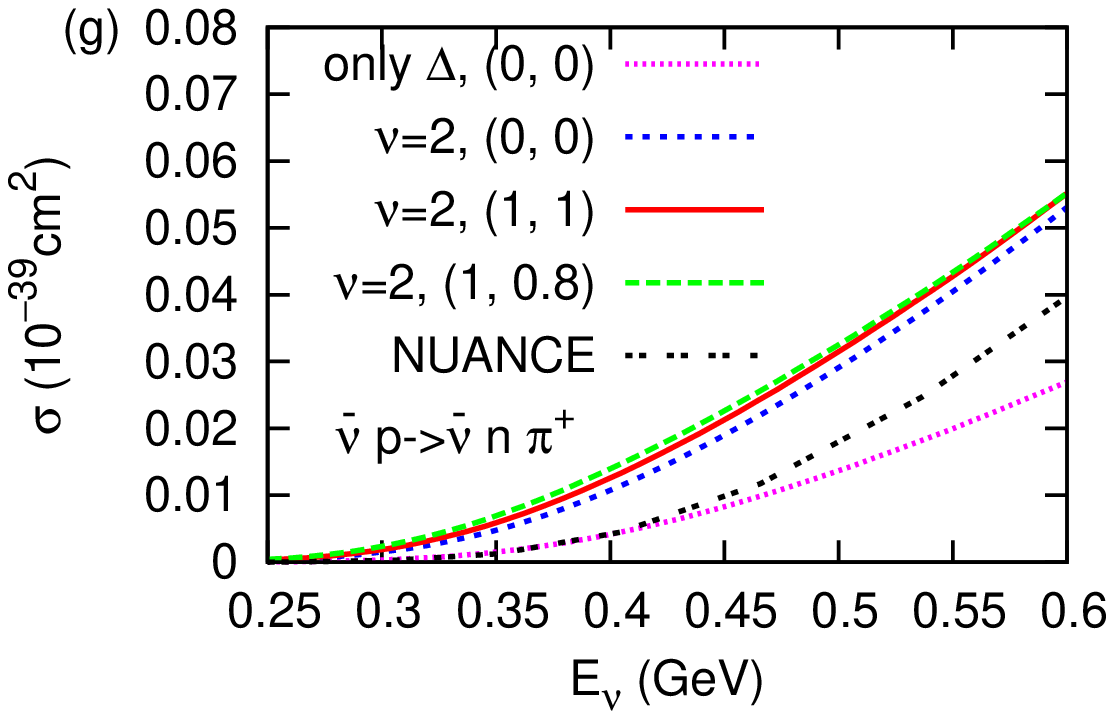}
\includegraphics[scale=0.66,angle=-90]{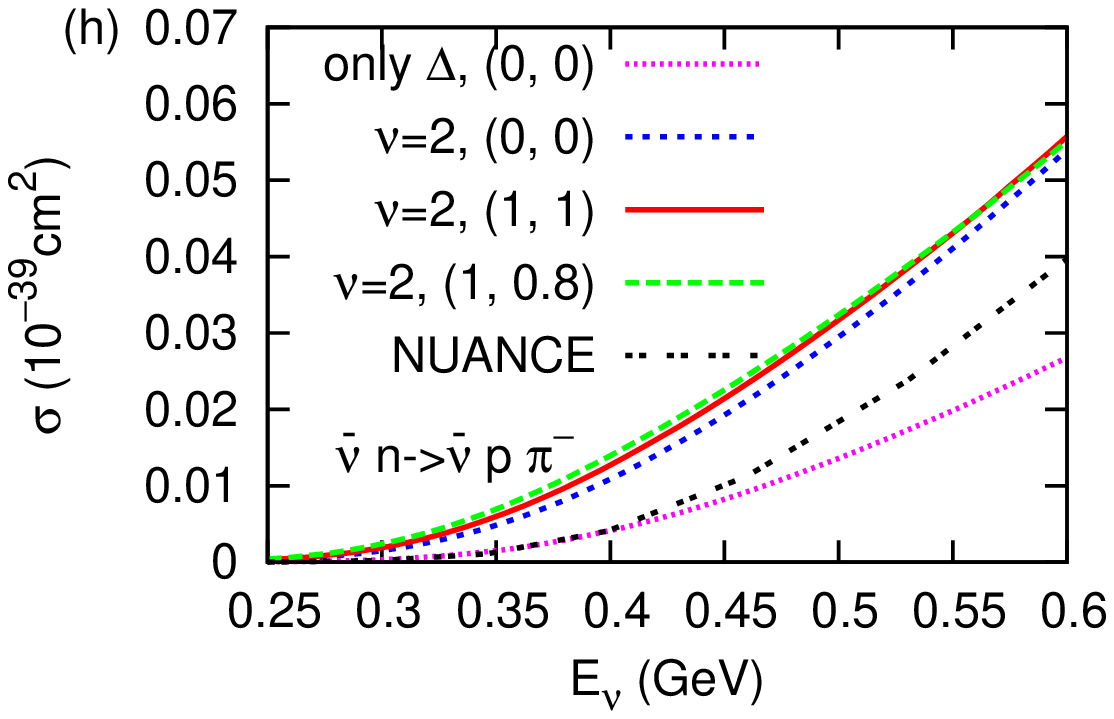}
\caption{(color online). Total cross section per proton or neutron for the NC pion production in neutrino-- and antineutrino--${}^{12}C$ scattering.
        }
\label{fig:p_ppion0_c12}
\end{center}
\end{figure}  

In Fig.~\ref{fig:p_ppion0_c12}, 
we show the total cross section for NC pion production from ${}^{12}C$. 
The categorization of the different calculations are 
the same as those for CC scattering.
Again the NUANCE output is close to our ``only 
$\Delta$'' calculation. Among the first three calculations in each channel, 
at fixed (anti)neutrino energy, $(1, \  0.8)$ 
gives a larger cross section than $(1, \  1)$ and $(1, \  1)$ gives a larger cross section than $(0, \  0)$. This is the same as in the CC production, 
which has been explained in terms of kinematics. Moreover, we can 
see how $\Delta$-dominance is violated in the NC case, as shown in
Tab.~\ref{tab:freeboundNC} (in which the labelings are the same as those in 
Tab.~\ref{tab:freeboundCC}, and free nucleon scattering results are 
from \cite{1stpaper}). The same is true for antineutrino--nucleus scattering.  
\begin{table}
 \centering
   \begin{tabular}{|c|c|c|c|c|c|c|} \hline
 $\sigma(10^{-39}\mathrm{cm}^{2})$        & only $\Delta$ (f)
           & $\nu=2$ (f)
           & Nonresonant (f)
           & Only $\Delta$ (b)
           & $\nu=2$ (b)
           & Nonresonant (b)   \\  \hline
$p,p\pi^{0}$ & $0.194$
           & $0.230$
           & $0.036$
           & $0.101$
           & $0.121$
           & $0.020$  \\   \hline
$n,n\pi^{0}$ & $0.194$
           & $0.234$
           & $0.040$
           & $0.101$
           & $0.123$
           & $0.022$  \\   \hline   
$n,p\pi^{-}$ & $0.089$
           & $0.149$
           & $0.060$
           & $0.045$
           & $0.082$
           & $0.037$  \\   \hline  
$p,n\pi^{+}$ & $0.089$
           & $0.155$
           & $0.066$
           & $0.045$
           & $0.088$
           & $0.043$  \\   \hline                                
   \end{tabular}
   \caption{Total cross sections averaged over number of proton or nucleon for NC pion production in neutrino--${}^{12}C$ scattering at $E_{\nu}=0.5$ GeV. See the text for detailed explanations. In the nuclear scattering,
   $r_{s}=r_{v}=1$.} \label{tab:freeboundNC}
\end{table}

\section{NC photon production} \label{sec:photonprod}

In this section, we study NC photon production from ${}^{12}C$.
The calculation is done in the same way as 
in pion production, except that the hadronic current in 
Eq.~(\ref{eqn:matrixelementNCcurrent}) is changed to the following:
\begin{eqnarray}
\langle J^{(had) \mu} \rangle \equiv \bra{N, \gamma} J^{(had) \mu} \ket{N} \ . 
\notag 
\end{eqnarray}   
The Feynman diagrams are the same as those in Fig.~\ref{fig:feynmanpionproduction} with the final $\pi$ line substituted by the final $\gamma$ line. 
See Ref.~\cite{1stpaper} for detailed discussion about them. Again we need to implement the change of the baryon 
spectrum when we apply the formula in \cite{1stpaper}, as we do 
in previous calculations.  
Because of built in symmetries in our model, 
conservation of the vector current is automatically satisfied, 
which is important for photon production.
The difference in the kinematic analysis, compared to that in pion 
production, is due to the zero mass of the photon. Moreover, we 
apply an energy cut on the photon energy in the laboratory frame, 
$E_{\gamma} \geqslant 0.15 \ \mathrm{GeV}$, motivated by the MiniBooNE's 
detector efficiency. This also eliminates the infrared singularity and simplifies 
the calculation.  

\begin{figure}
\begin{center}
\includegraphics[scale=0.66, angle=-90]{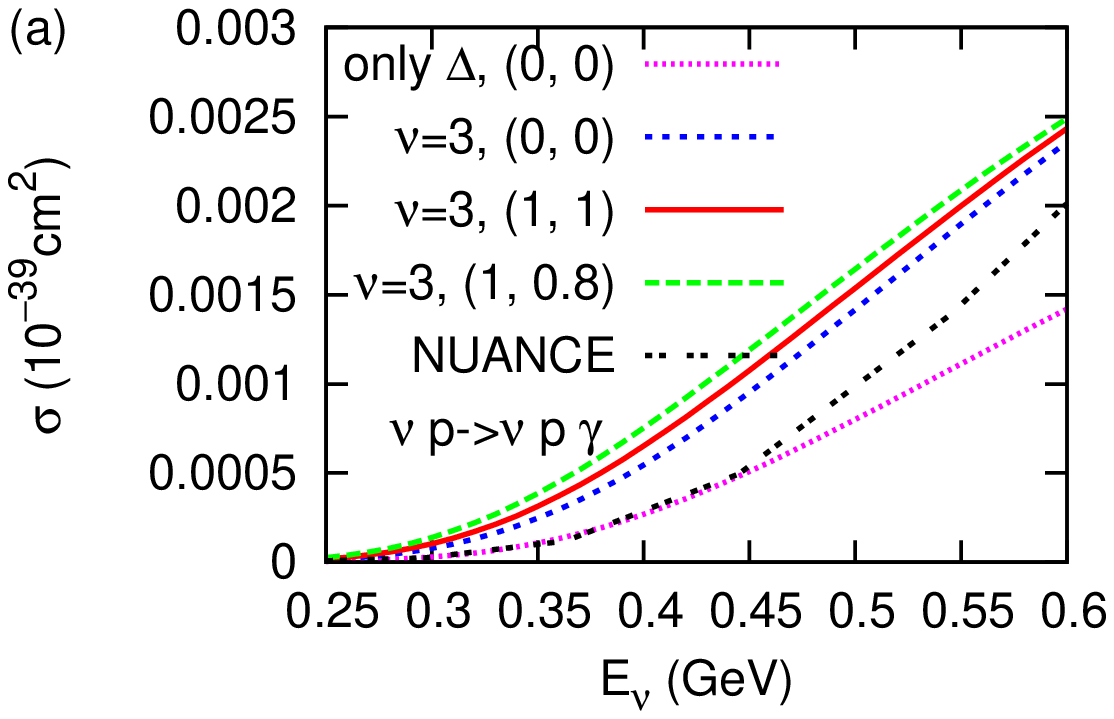}
\includegraphics[scale=0.66, angle=-90]{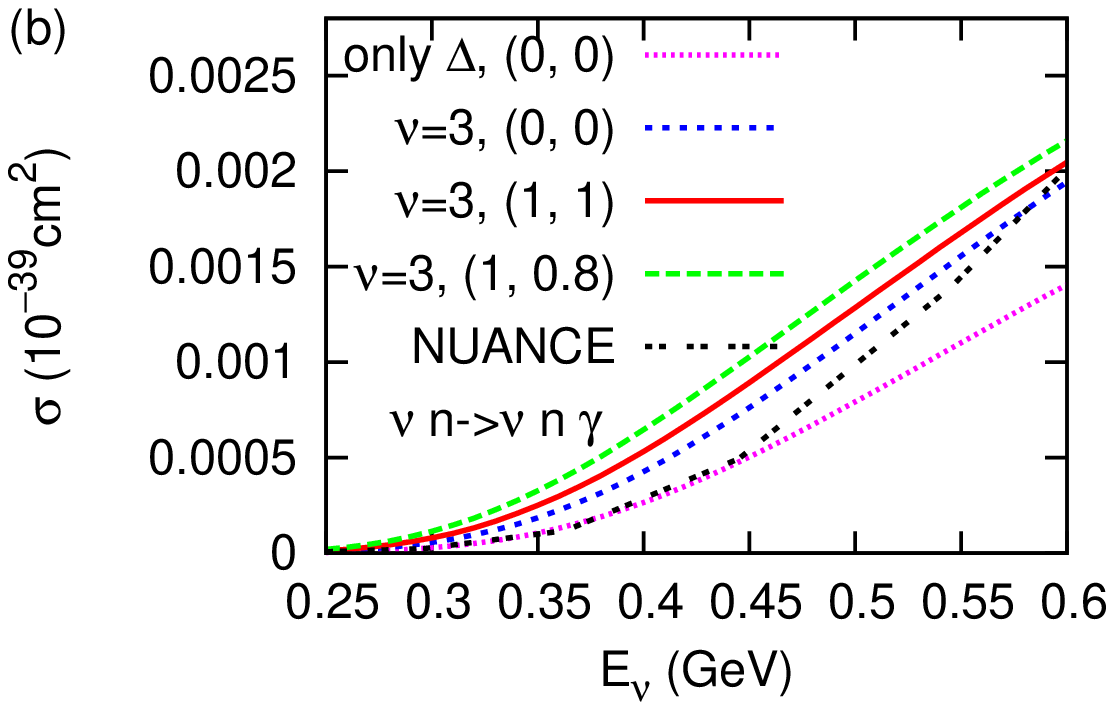}
\includegraphics[scale=0.66,angle=-90]{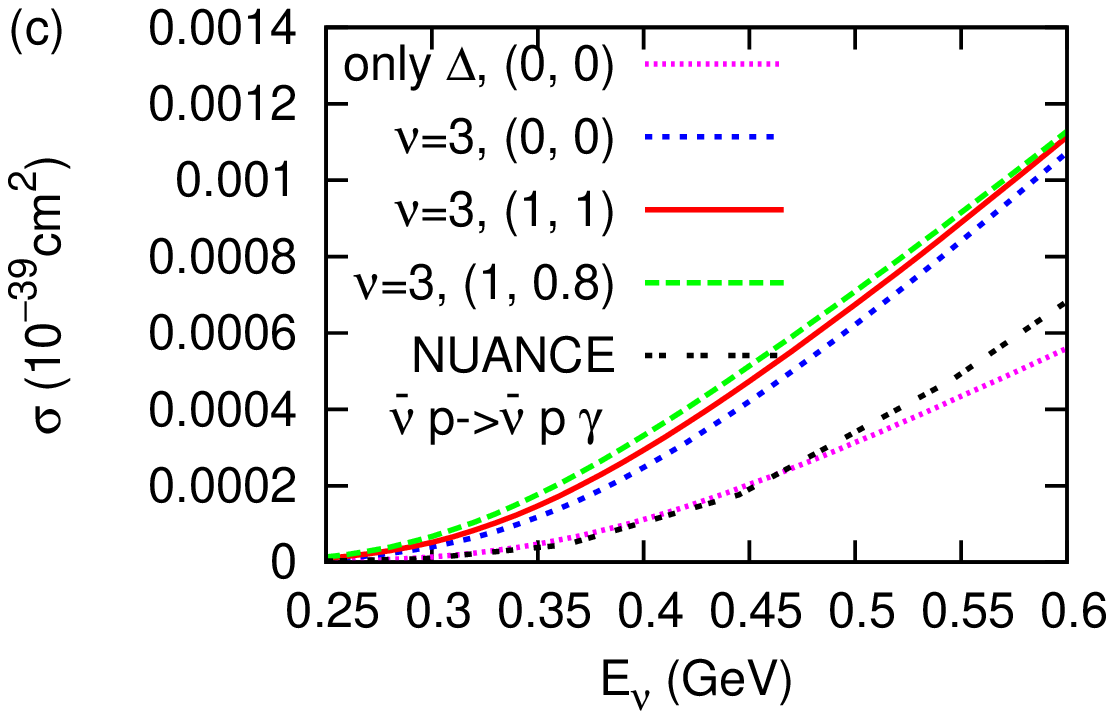}
\includegraphics[scale=0.66,angle=-90]{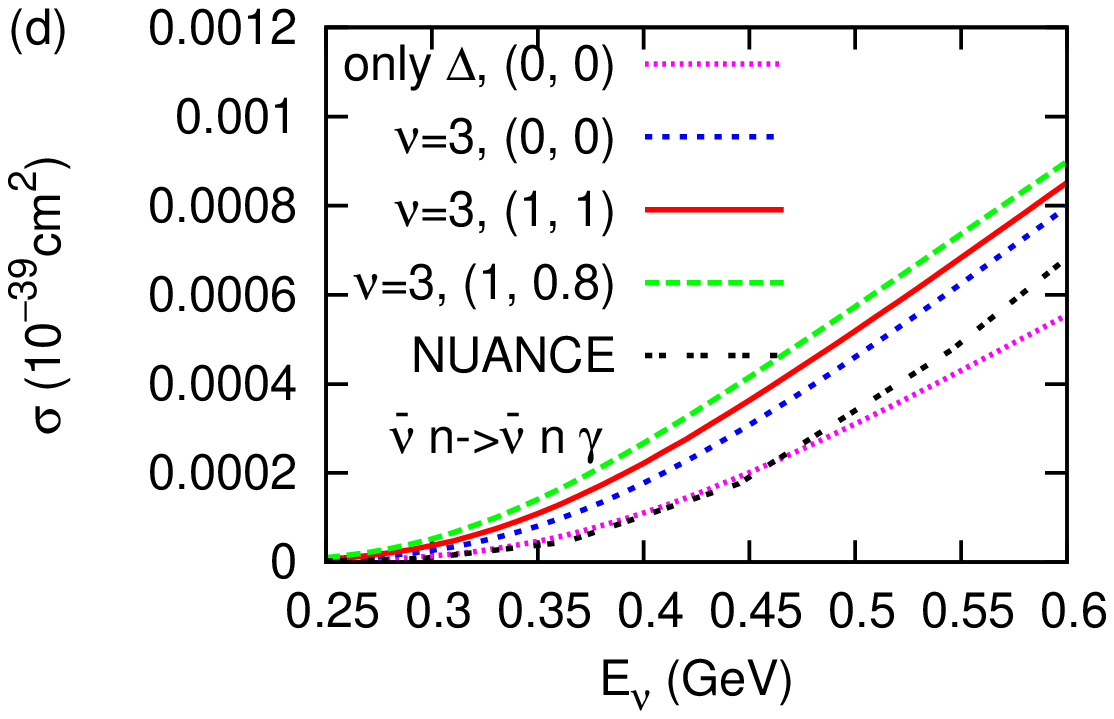}
\caption{(Color online) Total cross section per proton or neutron for the NC photon production 
          in the neutrino-- and antineutrino--${}^{12}C$ scatterings. Our calculation 
          is done with a photon energy cut 
          $E_{\gamma} \geqslant 0.15 \ \mathrm{GeV}$.
        }
\label{fig:p_pphoton_c12}
\end{center}
\end{figure}  

In Fig.~\ref{fig:p_pphoton_c12}, 
the total cross sections averaged over proton or neutron number 
are shown. Four different calculations are compared. The 
first ``only $\Delta$'' is the same as before. 
``$\nu=3$'' calculations include all the $\nu\leqslant 3 $ diagrams. 
It turns out no $\nu=2$ contact diagrams contribute, and there are only 
two $\nu=3$ contact vertices contributing (See Ref.~\cite{1stpaper} for details):
\[\frac{c_{1}}{M^{2}}\psibar{N}\ugamma{\mu} N\Tr \left(\widetilde{a}^{\nu} 
\psibar{F}^{(+)}_{\mu\nu}\right) \ , \ \frac{e_{1}}{M^{2}}\psibar{N}\ugamma{\mu} 
\widetilde{a}^{\nu}N \psibar{f}_{s\mu\nu} \ .\] 
As we have checked, the contributions of these two are small 
compared to those of the $\Delta$ and existing nonresonant diagrams, 
which should be expected according to the 
power-counting. Here, we have assumed their strength 
are due to both the $\omega$ and $\rho$ meson anomalous 
interaction vertices ($c_{1}=1.5$ and $e_{1}=0.8$) \cite{1stpaper,Hill10}.
Moreover for these calculations, changing $r_{s}$ and $r_{v}$ 
does not change the total cross section significantly, which is also 
observed in the differential cross section for pion electroproduction. 
In the three $\nu=3$ calculations for different channels, 
$(1, \  0.8)$ gives a bigger cross section than $(1, \  1)$ and $(1, \  1)$ is bigger 
than $(0, \  0)$. This pattern has been explained in pion production. 
We also see that the NUANCE
output is close to the ``only $\Delta$'' calculation and 
smaller than the full calculations, which should be expected from   
the comparison in pion production. 

In addition, in Tab.~\ref{tab:freeboundphoton}   
we show how the $\Delta$ significance changes from neutrino--nucleon scattering to 
neutrino--nuclus scattering (free nucleon scattering results are 
from \cite{1stpaper} with a change on photon energy 
cut: $E_{\gamma}\geqslant 0.15$ GeV): the $\Delta$ contribution is strongly reduced, 
and the nonresonant contribution is reduced less significantly. Since we have put 
a constraint on the minimum photon energy, the lower energy events are not 
included in the results and the Pauli blocking effect is not significant. That explains 
why the nonresonant contribution is not quite suppressed. And the reduction of the 
$\Delta$ contribution is mainly due to the broadening of its width. We also 
expect the Pauli blocking effect to be less significant with higher energy neutrinos. 
Furthermore, the same pattern about the reduction of cross sections 
happens in antineutrino scattering.
Based on Tab.~\ref{tab:freeboundphoton}, we need 
to include nonresonant contributions in photon production, 
as emphasized in pion production. 
\begin{table}
 \centering
   \begin{tabular}{|c|c|c|c|c|c|c|} \hline
 $\sigma(10^{-42}\mathrm{cm}^{2})$        & only $\Delta$ (f)
           & $\nu=1$ (f)
           & Nonresonant (f)
           & Only $\Delta$ (b)
           & $\nu=1$ (b)
           & Nonresonant (b)   \\  \hline
$p,p \gamma$ & $1.89$
           & $2.49$
           & $0.60$
           & $0.98$
           & $1.50$
           & $0.52$  \\   \hline
$n,n \gamma$ & $1.89$
           & $2.25$
           & $0.36$
           & $0.97$
           & $1.24$
           & $0.24$  \\   \hline                             
   \end{tabular}
   \caption{Total cross sections averaged over number of proton or nucleon for NC photon production in neutrino--${}^{12}C$ scattering at $E_{\nu}=0.5$ GeV. Here $E_{\gamma}\geqslant 0.15$ GeV for both types of scattering. 
   In the nuclear scattering, $r_{s}=r_{v}$=1.} \label{tab:freeboundphoton}
\end{table}

\section{summary} \label{sec:sum}

Neutrinoproduction of photons and pions from nuclei provides an important 
background in neutrino-oscillation experiment and must be 
understood quantitatively. Especially, we are interested in the possible role of 
NC photon production in the excess events seen in the MiniBooNE experiment at 
low \emph{reconstructed} neutrino energy. 
In Ref.~\cite{1stpaper}, we have 
calibrated our theory---QHD EFT with $\Delta$ introduced---by 
calculating photon and pion production from free nucleons up to 
$E_{\nu}=0.5 \ \mathrm{GeV}$.
In this work, the theory is applied to study the production from nuclei. 
Here we make use of the LFG approximation 
and Impulse Approximation, and include only one-body current
contributions. In the mean-field approximation of the nuclear 
ground state, the change of the baryon spectrum is represented by 
introducing an effective mass for baryons, which leads to the change of 
one-body currents in this calculation. The calculation for electron quasi-elastic 
scattering and electroproduction of pion serves as a benchmark 
for our approximation schemes. We then proceed to calculate 
the neutrinoproduction of pion and photon from ${}^{12}C$, and show the plots
for total cross section in every channel. 
First, we present calculations for pion production 
up to next-to-leading-order with 
different $r_{s}$ and $r_{v}$ parameters as constrained by the phenomenological 
study. It turns out that total cross sections are not very sensitive to changes 
of these parameters. Then in NC production of photon, although we show the result 
up to $\nu=3$ order, there are no $\nu=2$ contributions from contact terms, and as 
we have checked already the $\nu=3$ contributions due to $c_{1}$ and $e_{1}$, 
related to the so-called anomalous interactions, are tiny (the same has been shown 
for nucleon scattering in \cite{1stpaper}). 
Again, the total cross section of photon production is not sensitive to choice of 
different $r_{s}$ and $r_{v}$.
In all the plots, the $\Delta$ contributions are singled out and 
compared with the full calculations.
Moreover, we also compare our results with the output from NUANCE, and we find that
the NUANCE output is close to our ``only $\Delta$'' calculation with 
$(r_{s}=0, \ r_{v}=0)$ for both pion and photon production, which should 
be expected since the $\Delta$ dominates in NUANCE.    

In the calculation, the $\Delta$ dynamics 
in nuclei is a key component. The dynamics has been investigated 
in a nonrelativistic framework and also initiated in the QHD model. Parallel to the 
modification of the nucleon's spectrum, the $\Delta$-meson couplings (related to $r_{s}$ and $r_{v}$)
introduced in our theory dictates the real part of the $\Delta$ 
self-energy. The couplings are used to explain the 
S-L coupling of $\Delta$. Meanwhile the phenomenological 
result about S-L coupling based on nonrelativsic isobar-hole models 
puts an interesting constraint on the 
$\Delta$-meson coupling strengths, which is complementary to the constraints 
based on an EOS consideration. The $\Delta$ width is treated in a simplified way, as 
we take advantage of the existing result that shows an increase of the width 
due to the opening of other decay channels. In pion electroproduction, 
the pion-production (without FSI) result gives a correct prediction for the location of 
the $\Delta$-peak. We argue that this deficit is due to the absence of other channels. 
By adding contributions from two-body currents (from other relativistic studies) 
to our quasi-elastic and pion production (and turning off $\Delta$ broadening), 
we can explain the inclusive electron scattering strength. 
The investigation on $\Delta$ dynamics and two-body currents, 
which plays an important role in nuclear response and other problems, certainly needs to be pursued further in QHD EFT.

Moreover, because of the broadening of the $\Delta$ width, we expect that in 
both pion and photon productions, the $\Delta$ contribution is much less in 
nuclear scattering than in nucleon scattering. But the reduction of nonresonant 
contributions would be less at higher energies 
(beyond 0.5 GeV), because the Pauli blocking effect should be less important.
In Tabs.~\ref{tab:freeboundCC}, \ref{tab:freeboundNC}, and \ref{tab:freeboundphoton}, 
we have shown explicitly the cross sections at $E_{\nu}=0.5$ GeV due to 
$\Delta$ and nonresonant contributions in both neutrino--nucleon and neutrino--nucleus scattering. %
Although we see the reduction of nonresonant contributions for pion production 
in Tabs.~\ref{tab:freeboundCC} and \ref{tab:freeboundNC}, we see a smaller reduction 
for photon production in Tab.~\ref{tab:freeboundphoton}. This is consistent with 
the picture that the nonresonant contribution is reduced because of Pauli 
blocking. The same situation occurs in antineutrino scattering. 
This conclusion is important for future investigations of  
higher energy neutrino scattering, which may be relevant to MiniBooNE's excess 
event problem. 

Since our calculation is based on a QHD EFT Lagrangian with all the relevant
symmetries built in, conservation of vector current is manifest. 
This is crucial for photon production. Also partial conservation of the axial 
current is a necessary constraint in the problem. By using 
the mean-field approximation and the LFG model, 
these constraints are satisfied in a transparent way.

We are currently working on coherent pion and photon production from nuclei 
by applying this QHD EFT, which may also be relevant to the MiniBooNE low energy
excess event problem.

\acknowledgments
XZ would like to thank T. William Donnelly, Gerald T. Garvey, Joe Grange, Charles J. Horowitz, Teppei Katori, J. Timothy Londergan, William C. Louis, Rex Tayloe, and Geralyn Zeller for their valuable information, useful discussions, and important comments on the manuscript. This work was supported in part by the Department of Energy under Contract No.\ DE--FG02--87ER40365.

\appendix
\section{kinematics for quasi-elastic scattering} \label{app:quasikinematics}

The analysis of the kinematics is for scattering 
from nuclear matter, and can be easily generalized in the LFG model. 
The kinematic variables are shown in Fig.~\ref{fig:nucleiLabframeconfiguration}, 
and discussed following Eq.~(\ref{eqn:quasiTcrosssection}). 
From the mean-field theory in QHD EFT, we know that 
the leading order Hamiltonian gives rise to the nucleon
spectrum in nuclear matter as 
$p_{n}^{0} = g_{v} \langle V^{0} \rangle 
            +\sqrt{{\Mstar}^{2}
            +\vec{p}_{n}^{2}} $,   
$\Mstar \equiv M-g_{s} \langle \phi \rangle $.
Then we can define
${\starobject{p_{n}}}^{0} \equiv p_{n}^{0} 
                                -g_{v} \langle V^{0} \rangle
                          =\sqrt{{\Mstar}^{2}+\vec{p}_{n}^{2}}
                          =\sqrt{{\Mstar}^{2}
                                 +(\starobject{{\vec{p_{n}}}})^{2}}$.
This can be generalized from the laboratory frame to an arbitrary frame. 
In the LFG model, we consider each neighborhood inside the nucleus
as a homogeneous system; the field expectations, 
$\langle \phi(x) \rangle$ and $\langle V^{\mu}(x) \rangle$, 
are space-time dependent (and in the laboratory frame, they only depend 
on the space coordinate). In the following, we always work in the nuclear
laboratory frame. The covariance of our calculation is 
more transparent with the 
${\starobject{p_{n}}}^{\mu}$ variables than 
with $p_{n}^{\mu}$. For example energy momentum 
conservation is
$q+\starobject{p_{ni}}=\starobject{p_{nf}}$.  

Next we derive the formula for the total cross section. 
Suppose $M_{fi}$ is the covariant interaction amplitude 
between the probe and each individual nucleon with specific 
initial and final states. We have
\begin{eqnarray}
\sigma &=& \int dV \frac{1}{2 p_{li}^{0}} 
           \int \frac{d^{3}{\vec{p}_{nf}}^{\ast}}{(2\pi)^{3}2p_{nf}^{\ast 0}} 
           \frac{d^{3}\vec{p}_{lf}}{(2\pi)^{3} 2 p_{lf}^{0}} 
           \frac{d^{3}{\vec{p}_{ni}}^{\ast}}{(2\pi)^{3} 2 p_{ni}^{\ast 0}} 
           (2\pi)^{4} \delta^{4}(q+\starobject{p_{ni}}-\starobject{p_{nf}}) 
           \underset{s_{f},s_{i}}{\sum} \vert M_{fi} \vert^{2} \ . 
\end{eqnarray}
Pauli blocking leads to constraints 
on the integration of $\starobject{p_{ni}}$ and 
$\starobject{p_{nf}}$, i.e. 
$ \vert \vec{p}^{\ast}_{ni} \vert  \leqslant p_{F}$ 
and $\vert \vec{p}^{\ast}_{nf} \vert \geqslant p_{F}$. 
Here $p_{F}$ is the Fermi momentum related with the local density. 
The two constraints can be expressed by using factors 
$\theta[p_{F}^{2}+{\starobject{p_{ni}}}^{2}
-(\starobject{p_{ni}}\cdot V)^{2}/V^{2}]$ 
and 
$\theta[-p_{F}^{2}-{\starobject{p_{nf}}}^{2}
+(\starobject{p_{nf}}\cdot V)^{2}/V^{2}]$. 
In the following, we will not include them explicitly. 
We know that 
\begin{eqnarray}
\int \frac{d^{3} {\vec{p}_{ni}}^{\ast}}{2 p_{ni}^{\ast 0}} 
     \frac{d^{3} {\vec{p}_{nf}}^{\ast}}{2 p_{nf}^{\ast 0}} 
     \delta^{4}(q+\starobject{p_{ni}}-\starobject{p_{nf}})
 &=& \int d \phi_{\vec{p}_{ni}^{\ast}} 
          d p_{ni}^{\ast 0} 
          \frac{1}{4\modular{q}{}} 
          \quad \vert_{ \cos(\measuredangle\hat{q}\hat{p}_{ni}^{\ast})
                       =(2q^{0} p_{ni}^{\ast 0} +q^{2})
                        /(2\modular{q}{} \vert \vec{p}_{ni}^{\ast} \vert)
                      } \ . \notag 
\end{eqnarray}  
By using this, we have the total cross section as
\begin{eqnarray}
\sigma &=& \int dV \frac{1}{2p_{li}^{0}} 
           \int \frac{d^{3}\vec{p}_{lf}}{(2\pi)^{3} 2 p_{lf}^{0}} 
           \frac{d p_{ni}^{\ast0}}{\modular{q}{}} 
           \frac{d \phi_{\vec{p}_{ni}^{\ast}}}{16 \pi^{2}} 
           \underset{s_{f},s_{i}}{\sum} 
           \vert M_{fi} \vert^{2} \ . \label{eqn:xsectionquasi2} 
\end{eqnarray}
Meanwhile to make our phase space analysis simple, 
we can integrate over $d\modular{q}{}$ and $dq^{0}$:
\begin{eqnarray}
&& \sigma = \int \frac{dV}{(2\pi)^{4}} 
            d\phi_{\vec{p}_{ni}^{\ast}} 
            d p_{ni}^{\ast 0} 
            d q^{0}
            d \modular{q}{} 
            \frac{1}{16 p_{li}^{0} \modular{p}{li}} 
            \underset{s_{f},s_{i}}{\sum}
            \vert M_{fi} \vert^{2} \ . \label{eqn:xsectionquasi3}
\end{eqnarray}

Now, we need to calculate the boundary of the phase space 
in Eq.~(\ref{eqn:xsectionquasi3}). From the lepton 
kinematics, we can determine the boundary of 
$\modular{q}{}$ ($p_{li}^{0} \equiv E_{li}$):
\begin{eqnarray}
{\modular{q}{}}_{max} &=& \modular{p}{li} + \sqrt{E_{li}^{2}-M_{lf}^{2}}  
                         \ , \label{eqn:quasiphasespaceq1} \\[5pt]
{\modular{q}{}}_{min} &=& \modular{p}{li} - \sqrt{E_{li}^{2}-M_{lf}^{2}}  
                         \ . \label{eqn:quasiphasespaceq2} 
\end{eqnarray}
For a given $\modular{q}{}$, we have the following constraints 
based on the lepton kinematics:
\begin{eqnarray}
q^{0} &\leqslant& E_{li}-(E_{lf})_{min}
                                 =E_{li}-\sqrt{ (\modular{p}{li}-\modular{q}{})^{2}
                                               +M_{lf}^{2}
                                           \ ,    } \label{eqn:boundsonq01} \\[5pt]
q^{0} &\geqslant& E_{li}-(E_{lf})_{max} = 0 \ . \label{eqn:boundsonq02}
\end{eqnarray}
However, there are further constraints on $q^{0}$ for a 
given $\modular{q}{}$ due to the hadron 
kinematics. 
For a given set of $q^{0}$, $\modular{q}{}$,  
$\cos(\measuredangle\hat{q}\hat{p}_{ni}^{\ast})=(2q^{0} p_{ni}^{\ast 0} 
+q^{2})/(2\modular{q}{} \vert \vec{p}_{ni}^{\ast} \vert )$ 
has to be physical. This requires
\begin{eqnarray}
&& \vert \cos(\measuredangle\hat{q}\hat{p}_{ni}^{\ast}) \vert 
   \leqslant 1 \notag \\[5pt]
\Longleftarrow 
&& \vert \vec{p}_{ni}^{\ast}\vert 
   \geqslant \left| \frac{\modular{q}{}}{2}
                   -\frac{q^{0}}{2}\sqrt{1-\frac{4{\Mstar}^{2}}{q^{2}}} 
            \right| 
   \equiv p^{-} \ . \label{eqn:quasiphasespacepni1} 
\end{eqnarray}
Eq.~(\ref{eqn:quasiphasespacepni1}) gives a lower bound 
of $\vert \vec{p}_{ni}^{\ast}\vert$ which is also required to be below the Fermi surface: 
$\vert \vec{p}_{ni}^{\ast}\vert \leqslant p_{F}$. Combining $p_{F} \geqslant p^{-}$ 
and the constraints in Eqs.~(\ref{eqn:boundsonq01}) and (\ref{eqn:boundsonq02}), 
we find 
\begin{eqnarray}
q^{0}_{min} &=& \max\left[\sqrt{(\modular{q}{}-p_{F})^{2}
                                              +{\Mstar}^{2}}-E_{F}
                                    , 0\right] \ , \label{eqn:quasiphasespaceq01} \\[5pt]
q^{0}_{max} &=& \min \left[\sqrt{(\modular{q}{}+p_{F})^{2}+{\Mstar}^{2}}
                                     -E_{F},
                                     E_{li}-\sqrt{(\modular{p}{li}-\modular{q}{})^{2}
                                                  +M_{lf}^{2}}
                                    \right] \ . \label{eqn:quasiphasespaceq02} 
\end{eqnarray}
Moreover, the constraint 
$\vert \vec{p}_{nf}^{\ast} \vert \geqslant p_{F}$ is 
not present in the former discussion, but is taken care of 
in the numerical calculation.  

\section{kinematics for pion production} \label{app:pionprodkinematics}
The kinematic variables are defined in 
Fig.~\ref{fig:nucleipionprodLabframe} in the laboratory frame. 
Except for the $\pi$ momentum $k_{\pi}$, all the others are 
defined in Appendix~\ref{app:quasikinematics}. The variables 
defined in other frames will be mentioned explicitly. First we have
\begin{eqnarray}
\sigma &=& \int dV \frac{1}{2 p_{li}^{0}}
           \int \frac{d^{3}{\vec{p}_{nf}}^{\ast}}
                     {(2\pi)^{3} 2 p_{nf}^{\ast 0}} 
                \frac{d^{3}\vec{k}_{\pi}}
                     {(2\pi)^{3} 2 k_{\pi}^{0}}
                \frac{d^{3}\vec{p}_{lf}}{(2\pi)^{3} 2 p_{lf}^{0}} 
                \frac{d^{3}{\vec{p}_{ni}}^{\ast}}
                     {(2\pi)^{3} 2 p_{ni}^{\ast 0}} \notag \\[5pt]
        && \times (2\pi)^{4} 
                  \delta^{4} (q+\starobject{p_{ni}}-\starobject{p_{nf}}-k_{\pi}) 
                  \underset{s_{f},s_{i}}{\sum} 
                  \vert M_{fi} \vert^{2} \ . \notag
\end{eqnarray}
The constraints on 
$\vec{p}_{ni}^{\ast}$ and $\vec{p^{\ast}}_{nf}$, i. e. 
$ \vert \vec{p}^{\ast}_{ni} \vert  \leqslant p_{F}$ 
and $\vert \vec{p}^{\ast}_{nf} \vert \geqslant p_{F}$, are always 
implicit in the formula.

One way to think about the phase space as follows: 
Given specific values for $q$ and $p_{ni}^{\ast}$, the final pion 
and nucleon invariant mass $M_{\pi n}$ 
are fixed, and then the degrees of freedom in the 
isobaric frame (final pion and nucleon's center-of-mass frame) 
is the angle of $\vec{k}_{I\pi}$, i.e. 
$\Omega_{\vec{k}_{I\pi}}$. So we have
\begin{eqnarray}
&& \int \frac{d^{3}{\vec{p}_{nf}}^{\ast}}
             {(2\pi)^{3} 2 p_{nf}^{\ast 0}} 
        \frac{d^{3}\vec{k}_{\pi}}
             {(2\pi)^{3} 2 k_{\pi}^{0}} 
        (2\pi)^{4} \delta^{4}
        (q+\starobject{p_{ni}}-\starobject{p_{nf}}-k_{\pi}) \notag \\[5pt]
&=& \int dM_{n\pi} d\Omega_{\vec{k}_{I\pi}}  
         \frac{1}{(2\pi)^{2}} \frac{\modular{k}{I\pi}}{2} 
         \delta[(q+p_{ni}^{\ast})^{2}-M_{\pi n}^{2}] 
         \ . \notag 
\end{eqnarray}
In the above, we have made use of the following identities:
\begin{eqnarray}
M_{\pi n} &\equiv& \sqrt{{\Mstar}^{2}+\modular{k}{I\pi}^{2}}
                  +\sqrt{{M_{\pi}}^{2}+\modular{k}{I\pi}^{2}} 
                  \ , \notag \\[5pt]
E_{I\pi}  &=& \frac{M_{\pi n}^{2}-{\Mstar}^{2}+M_{\pi}^{2}}
                   {2M_{\pi n}}   \qquad 
              E_{Inf} = M_{\pi n}-E_{I \pi} 
             \ , \notag \\[5pt]
\frac{d E_{I \pi}}{dM_{\pi n}} 
          &=& \frac{E_{Inf}}{M_{\pi n}} \ . \notag 
\end{eqnarray}
Then analogous to the analysis in 
the quasi-elastic scattering case, we have
\begin{eqnarray}
&&  \int \frac{d^{3}{\vec{p}_{ni}}^{\ast}}
              {(2\pi)^{3} 2 p_{ni}^{\ast0}}
         \delta[(q+p_{ni}^{\ast})^{2}-M_{\pi n}^{2}]
    \notag \\[5pt]
&=& \int \frac{d p_{ni}^{\ast 0} d \phi_{\vec{p}_{ni}^{\ast}}}
              {4 \modular{q}{}(2\pi)^{3}} 
         \vert_{ \cos(\angle\hat{q}\hat{p}_{ni}^{\ast}) 
                = (2q^{0}p_{ni}^{\ast 0} 
                  + q^{2}+{\Mstar}^{2} 
                  - M_{\pi n}^{2})/2\modular{q}{} 
                    \vert \vec{p}_{ni}^{\ast} \vert
               } \ . \notag 
\end{eqnarray}
So finally:
\begin{eqnarray}
\sigma &=& \int dV \frac{1}{2 p_{li}^{0}} 
           \int \frac{d^{3}\vec{p}_{lf}}
                     {(2\pi)^{3} 2 p_{lf}^{0}} 
           \int dM_{n\pi} d\Omega_{\vec{k}_{I\pi}} 
                dp_{ni}^{\ast 0} d\phi_{\vec{p}_{ni}^{\ast}}
                \frac{1}{(2\pi)^{5}} 
                \frac{\modular{k}{I\pi}}{8\modular{q}{}} 
                \underset{s_{f},s_{i}}{\sum} 
                \vert M_{fi} \vert^{2} 
           \notag \\[5pt]
       &=& \int dV dq^{0} d\modular{q}{} dM_{\pi n} 
                dp_{ni}^{\ast 0} d\phi_{\vec{p}_{ni}^{\ast}} 
                d\Omega_{\vec{k}_{I\pi}} 
                \frac{1}{(2\pi)^{7}} 
                \frac{\modular{k}{I\pi}}{32 (p_{li}^{0})^{2}} 
                \underset{s_{f},s_{i}}{\sum} 
                \vert M_{fi} \vert^{2} 
                \ . \notag
\end{eqnarray}

Next, we need to determine the boundary of phase space in 
terms of these variables. 
First, it is clear that no constraint needs to be applied to 
$\Omega_{\vec{k}_{I\pi}}$ and $\phi_{\vec{p}_{ni}^{\ast}}$.
Second, for a given set of $\vec{q}, q^{0}$, and $M_{\pi n}$,
to make sure 
$\vert \cos(\angle\hat{q}\hat{p}_{ni}^{\ast})\vert \leqslant 1 $,
there is a constraint on $p_{ni}^{\ast 0}$ besides 
$p_{ni}^{\ast 0} \leqslant E_{F}$: 
\begin{eqnarray}
&& -2 \modular{q}{} \vert \vec{p}_{ni}^{\ast} \vert -q^{2} 
   \leqslant 2q^{0} p_{ni}^{\ast 0}
            +{\Mstar}^{2} 
            -M_{\pi n}^{2} 
   \leqslant 2\modular{q}{} \vert \vec{p}_{ni}^{\ast} \vert - q^{2} 
   \ , \notag \\[5pt] 
& \Longleftrightarrow &  
\begin{cases}
   (\vert \vec{p}_{ni}^{\ast} \vert 
    +\frac{\lambda+1}{2} \modular{q}{}
   )^{2} 
    +\frac{{\Mstar}^{2} (q^{0})^{2}}{q^{2}} 
    -\frac{(\lambda+1)^{2}}{4} (q^{0})^{2} 
    \geqslant 0  \ ,
    \qquad 
    \lambda \equiv \frac{M_{\pi n}^{2}-{\Mstar}^{2}}
                        {-q^{2}}
    \ ; \notag \\[5pt]
  q^{0} \sqrt{ \vert \vec{p}_{ni}^{\ast}\vert^{2}
                            +{\Mstar}^{2}
                           }
               +\modular{q}{} \vert \vec{p}_{ni}^{\ast} \vert 
               \geqslant -\frac{q^{2}}{2} 
                         +\frac{M_{\pi n}^{2}-{\Mstar}^{2}}{2} 
              \ ,  \notag 
\end{cases} \notag \\[5pt] 
&\Longleftrightarrow &  \vert \vec{p}_{ni}^{\ast} \vert 
                        \geqslant p^{-} 
                                  \equiv 
                                  \left| \frac{\lambda+1}{2} 
                                         \modular{q}{} 
                                        -\frac{q^{0}}{2} 
                                         \sqrt{(\lambda+1)^{2}
                                        -\frac{4{\Mstar}^{2}}{q^{2}}}
                                 \right|  \ . 
                        \label{eqn:boundsonEni}
\end{eqnarray}
Third, for a given set of $\vec{q}, q^{0}$, there is a constraint 
on $M_{\pi n}$, such that $p^{-} \leqslant p_{F}$. 
From $M_{\pi n}^{2}\equiv q^{2}+{\Mstar}^{2} +2q^{0} p_{ni}^{\ast 0} -2 \modular{q}{} \vert 
\vec{p}_{ni}^{\ast}\vert \cos(\angle\hat{q}\hat{p}_{ni}^{\ast})$, 
we have
\begin{eqnarray}
&& q^{2} + {\Mstar}^{2} + 2q^{0} E_{F} - 2\modular{q}{} p_{F} 
   \leqslant M_{\pi n}^{2} 
   \leqslant q^{2}+{\Mstar}^{2} + 2q^{0} E_{F} +2 \modular{q}{} p_{F} 
   \ . \notag 
\end{eqnarray}
And to open the pion production threshold, we need
\begin{eqnarray}
(M_{\pi}+\Mstar)^{2} \leqslant M_{\pi n}^{2} \ , \notag
\end{eqnarray}
So, we have
\begin{eqnarray}
\max((M_{\pi}+\Mstar)^{2},  q^{2}+{\Mstar}^{2} 
                           +2q^{0} E_{F} -2 \modular{q}{} p_{F}
    ) && \notag \\[5pt] 
\leqslant M_{\pi n}^{2} \leqslant q^{2} 
      &+& {\Mstar}^{2} + 2q^{0} E_{F} +2 \modular{q}{} p_{F}    
          \ . \label{eqn:boundsonMpin}
\end{eqnarray} 
Fourth, for a given $\modular{q}{}$, there is a constraint on $q^{0}$ 
such that $(M_{\pi n})_{ min} \leqslant (M_{\pi n})_{ max}$. 
We have
\begin{eqnarray}
&& q^{2}+{\Mstar}^{2} +2q^{0} E_{F} +2 \modular{q}{} p_{F} 
   \geqslant (M_{\pi}+\Mstar)^{2} \notag \\[5pt]
\Longleftrightarrow 
&& q^{0} \geqslant \max( \sqrt{ (M_{\pi}+\Mstar)^{2}
                               +(\modular{q}{}-p_{F})^{2}
                              }
                        -E_{F}, 0
                       )  \ . 
\label{eqn:pionprodboundsonq01}
\end{eqnarray}
However, there are further constraints on $q^{0}$ due to the 
lepton kinematics, which has been shown in Eqs.~(\ref{eqn:boundsonq01}) 
and (\ref{eqn:boundsonq02}). Together with Eq.~(\ref{eqn:pionprodboundsonq01}), 
we have the boundary of $q^{0}$:
\begin{eqnarray}
q^{0}_{max} &=&  E_{li}-\sqrt{ (\modular{p}{li} -\modular{q}{})^{2}
                              + M_{lf}^{2}
                             }  \ , 
            \label{eqn:pionprodboundsonq02} \\[5pt]
q^{0}_{min} &=& \max( \sqrt{ (M_{\pi}+\Mstar)^{2}
                            +(\modular{q}{}-p_{F})^{2}
                           }
                      -E_{F}, 0
                    )  \ . 
            \label{eqn:pionprodboundsonq03}
\end{eqnarray}
Eqs.~(\ref{eqn:quasiphasespaceq1}) and (\ref{eqn:quasiphasespaceq2}) 
give constraints on $\modular{q}{}$. And there are further 
constraints due to hadron kinematics. We have to 
make sure that $q^{0}_{max} \geqslant q^{0}_{min}$. But it is complicated 
to obtain an analytic expression for $\modular{q}{}$ based on this constraint. 
In the numerical calculations, 
we made use of another boundary by assuming 
a static nucleon in vacuum. Then, we can 
say the allowed region of $\modular{q}{}$ is always inside the previous region. 
Solving 
$q^{0}_{max} \geqslant q^{0}_{min}$ with $p_{F}=0$ and $\Mstar=M$ gives us:
\begin{eqnarray}
\modular{q}{} &\leqslant& \frac{ \beta_{A} \frac{ M_{A}^{2}
                                                 +(M_{\pi}+M)^{2}
                                                 -M_{lf}^{2}
                                                }{E_{li}+M}
                                + \sqrt{\Delta}
                               }{2(1-\beta_{A}^{2})} 
                          \ ,   \label{eqn:boundsonq1} \\[5pt]
\modular{q}{} &\geqslant& \frac{ \beta_{A} \frac{ M_{A}^{2}
                                                 +(M_{\pi}+M)^{2}
                                                 -M_{lf}^{2}
                                                 }{E_{li}+M}
                                -\sqrt{\Delta}
                               }{2(1-\beta_{A}^{2})} 
                        \ .    \label{eqn:boundsonq2}
\end{eqnarray}
In the above, 
\begin{eqnarray}
\beta_{A} &=& \frac{E_{li}}{E_{li}+M} \ , \notag \\[5pt]
M_{A}     &=& \sqrt{(E_{li}+M)^{2}-E_{li}^{2}} \ , \notag \\[5pt]
\Delta    &=& \frac{[M_{A}^{2}+(M_{\pi}+M)^{2}-M_{lf}^{2}]^{2}}
                   {(E_{li}+M)^{2}} 
             -4(1-\beta_{A}^{2})(M_{\pi}+M)^{2} \ . \notag
\end{eqnarray}
So, Eqs.~(\ref{eqn:quasiphasespaceq1}), (\ref{eqn:quasiphasespaceq2}), 
(\ref{eqn:boundsonq1}) and (\ref{eqn:boundsonq2}) are the bounds 
used in the numerical calculations. And to map out the physical 
region, we simply try and check. 
It is also complicated to determine an analytic 
expression for the threshold value of $E_{li}$ 
in the LFG model. 
However, it is simpler to work out the value 
for pion production off a static nucleon, which is
\begin{eqnarray}
E_{li} \geqslant \frac{(M_{\pi}+M+M_{lf})^{2}-M^{2}}{2M} \ .  \notag 
\end{eqnarray}
So, the difference between the true threshold and the value 
calculated above is essentially the binding energy.

\end{document}